\documentclass[10pt,letterpaper]{article}
\usepackage[top=0.85in,left=1.2in,right=1.5in,bottom=1in]{geometry}
\usepackage[utf8]{inputenc}

\usepackage{cite}

\usepackage{nameref,hyperref}
\usepackage{amsmath,amssymb,bm} 
\usepackage{capt-of}
\usepackage{graphicx,times}
\usepackage{float}
\usepackage{color}
\usepackage{array}
\usepackage{multirow}
\usepackage{multicol}
\usepackage{fixltx2e}
\usepackage{verbatim}
\usepackage{subcaption}

\usepackage{microtype}
\DisableLigatures[f]{encoding = *, family = * }

\usepackage{changepage}

\usepackage[aboveskip=1pt,labelfont=bf,labelsep=period,singlelinecheck=off]{caption}

\makeatletter
\renewcommand{\@biblabel}[1]{\quad#1.}
\makeatother

\usepackage{lastpage,fancyhdr,graphicx}
\usepackage{epstopdf}
\fancyheadoffset[L]{1.25in}
\fancyfootoffset[L]{2.25in}

\usepackage{color}

\definecolor{Gray}{gray}{.25}

\usepackage{graphicx}

\usepackage{sidecap}

\usepackage{wrapfig}
\usepackage[pscoord]{eso-pic}
\usepackage[fulladjust]{marginnote}
\reversemarginpar

\begin{document}
\vspace*{0.25in}

\begin{flushleft}
{\Large
\textbf\newline{Compressed Convolutional LSTM: An Efficient Deep Learning Framework to Model High Fidelity 3D Turbulence}
}
\newline
\\
\textit{Arvind T. Mohan} \textsuperscript{1,2},
Don Daniel \textsuperscript{2},
Michael Chertkov\textsuperscript{2,3} \&
Daniel Livescu \textsuperscript{2} 
\\
\bigskip
\textsuperscript{1} \textrm{\small Center for Nonlinear Studies, Los Alamos National Laboratory}
\\
\textsuperscript{2} \textrm{\small Los Alamos National Laboratory, Los Alamos, NM 87545, USA}
\\
\textsuperscript{3} \textrm{\small Department of Applied Mathematics, University of Arizona, Tucson, AZ 85721, USA}
\\
\bigskip
\textrm{\small Corresponding author: \textit{arvindm@lanl.gov}}

\end{flushleft}

\section*{Abstract}
High-fidelity modeling of turbulent flows is one of the major challenges in computational physics, with diverse applications in engineering, earth sciences and astrophysics, among many others. The rising popularity of high-fidelity computational fluid dynamics (CFD) techniques like direct numerical simulation (DNS) and large eddy simulation (LES) have made significant inroads into the problem. However, they remain out of reach for many practical three-dimensional flows characterized by extremely large domains and transient phenomena. Therefore designing efficient and accurate data-driven generative approaches to model turbulence is a necessity. We propose a novel training approach for dimensionality reduction and spatio-temporal modeling of the three-dimensional dynamics of turbulence using a combination of Convolutional autoencoder and the Convolutional LSTM neural networks. The quality of the emulated turbulent fields is assessed with rigorous physics-based statistical tests, instead of visual assessments. The results show significant promise in the training methodology to generate physically consistent turbulent flows at a small fraction of the computing resources required for DNS.

\section{Introduction}
\label{intro}
Several research problems in seemingly disparate fields such as socio-economics, infrastructure networks,  physical and natural sciences etc., have a common thread: The data from these systems consist of multiple features varying in both space and time. i.e. they exhibit strong \textit{spatio-temporal} dynamics. In addition, many of these systems are high dimensional with several thousands/million of degrees of freedom, making them exceptionally complex to study and theoretically model due to the challenges in standard mathematical, statistical tools to handle them. Therefore, such systems are often measured or modeled computationally, often producing vast amounts of high dimensional data. However, in many cases it is impossible to experimentally measure the full spatio-temporal system state, for applications like earth sciences and remote sensing. Additionally, it may also be prohibitively expensive to computationally model the system at high fidelity with governing equations. In some applications, the terms in governing equations may not be completely known; leading to errors in computational models. As a result, there is an urgent need to leverage existing data and observations to identify important patterns and system dynamics, which can enable low cost and high fidelity models, with better theoretical insight into the physics of the system. \textit{Data driven modeling} of such systems is a fruitful, yet recent area of investigation, with initial successes demonstrated by Pathak et. al~\cite{pathak2018model,pathak2017using} for low dimensional chaotic systems. One such system is turbulence in fluids, which is extremely high-dimensional, non-linear and chaotic, with multiple applications in engineering, earth sciences, climate and biological systems. This is the focus of the present work, though the ideas discussed herein may be extendable to other domains.

In the recent years, Deep learning (DL) techniques powered by neural networks have established themselves as state-of-the art for data driven models, with successes in myriad applications in information technology, socio-economics and of late, in science. There has been a surge of interest in DL applications to fluid mechanics, specifically to computational fluid dynamics (CFD) of turbulent flows. Several advancements in DL and classical machine learning techniques for CFD are focused in the area of turbulence modeling for Reynolds averaged Navier Stokes (RANS) and subgrid modeling for Large Eddy Simulation (LES). In these approaches, the turbulent scales are intelligently parameterized for a class of flows by learning them from high fidelity Direct Numerical Simulation (DNS) databases. Several of these approaches have augmened existing turbulence models with machine learning~\cite{wu2018physics,wang2017comprehensive}, exploited inverse modeling~\cite{tracey2015machine,singh2017machine} with classical machine learning algorithms, while and others have utilized neural networks to learn reynolds stress closures~\cite{maulik2019subgrid,ling2016reynolds}, thereby reducing computational costs of RANS/LES and increasing accuracy.

However, there exist applications where solely parameterizations of turbulence - or even CFD - may not be ideal approaches. An example is the generation of inflow turbulence respecting specific boundary/initial conditions, which is crucial to accurately initiate high fidelity CFD simulations~\cite{klein2003digital,di2006synthetic} and there have been several efforts to model synthetic turbulence~\cite{juneja1994synthetic,jarrin2006synthetic}. Several examples can be found in geophysical modeling, where the sheer diversity of multiphysics phenomena and range of turbulence scales makes high fidelity multi-scale Navier Stokes solutions computationally intractable, even for small geographic regions, requiring modeling of various unresolved scales~\cite{duan2007stochastic,o2008statistical}. Applications in astrophysics are characterized by extremely large domains and transient phenomena which require mesh and temporal resolution often outside the realm of available computing resources~\cite{schmidt2011fluid}.
In all these examples, there is an urgent need to model the spatio-temporal dynamics of turbulence which are physically consistent, at a low computational cost. The challenge then boils down to \textit{learning the underlying attractor of the dynamical system} (turbulence), following which the learned model of this attractor can be used to generate subsequent spatio-temporal realizations obeying the governing laws of the dynamical system. As a result, there has been sustained interest in reduced order modeling of turbulence, where the standard Navier-Stokes PDEs are distilled to simplified, computationally cheap forms to model a variety of flow regimes with Galerkin projection based methods~\cite{rempfer2000low,noack2005need,carlberg2011efficient}. Other novel approaches use sparse coding~\cite{deshmukh2016model,sargsyan2015nonlinear}, cluster based~\cite{kaiser2014cluster} and network based~\cite{nair2015network} approaches. 

Simultaneously, several advances in the recent decade have occurred in using stacked neural networks i.e. Deep learning for image classification and temporal modeling, which are often the nature of problems in the information industry. As a result, progress in DL for modeling spatio-temporal complexity in high dimensions has not evolved at the same rate. However, its ubiquity in real-world applications has caught the attention of the community in the recent years, with significant advances being made in generative models~\cite{goodfellow2014generative,xingjian2015convolutional} for  image generation and video classification/generation.
Much of the interest has been from the computer graphics and animation community, with several recent works focusing on realistic fluid animations~\cite{chu2017data}, splash animation~\cite{um2018liquid} and droplet dynamics~\cite{mukherjee2018neuraldrop}. Many of these efforts demonstrated potential of deep learning methods such as Convolutional and Generative Adversarial networks to handle such complex datasets. As a result of focus on animation and graphics, modeling of large, highly mulitiscale phenomena critical for engineering applications have not been explored with rigorous analysis of the predicted physics. Recent efforts from physics community have been centered over reservoir computing~\cite{zimmermann2018observing,lu2017reservoir,pathak2018hybrid}, and for the first time the approach has recently been shown to model a chaotic system such as 1-dimensional the Kuramoto-Sivashinsky equation~\cite{pathak2018model}. There have been more efforts on using prior knowledge of the system and learning algorithms to model the attractor of the dynamical system, with promising results~\cite{pathak2018hybrid}, including efforts using generative deep learning models~\cite{farimani2017deep}. At this juncture, there have been scarce works on evaluating the capability of deep learning approaches to model multiscale turbulence. In particular, there is a need to develop compute efficient deep learning approaches for three-dimensional turbulence, due to its ubiquity in practical applications. In addition, it is important to study the mechanics of the ``black box" neural networks from the perspective of turbulence physics, to demystify it for future applications. 

A major obstacle in extending neural networks to such 3D physics problems is the extremely high-dimensional nature of the data. To illustrate this, we present a contrived example of training LSTM (Long Short Term Memory) neural networks, which are specialized recurrent networks excelling in modeling temporal or sequential dynamics~\cite{sutskever2014sequence}, on a high-dimensional system. Consider a simplistic two-dimensional fluid flow comprised of a few thousand time-varying data points. The naive approach of utilizing LSTMs for this problem is to train them for each data point in the 2D surface. This quickly makes the approach computationally prohibitive since the number of models trained scales linearly with the number of data points in the domain. Furthermore, this approach is prone to failure since the network has no direct way of capturing spatial correlations between data points. The lack of `communication' between spatial data points can lead to computationally intractable and erroneous models.

This work demonstrates an innovative approach of exploiting recent advancements in Convolutional neural networks, LSTMs and neural network autoencoders with the goal of a) Dimensionality reduction of high-dimensional turbulence datasets, and b) Learning the attractor of the governing dynamical system to model its spatio-temporal dynamics at a low computational cost. Our specific focus is on turbulence in fluids, though the we hope the ideas here are transferable to other multiscale phenomena. To the best of our knowledge, this is the first attempt to directly model three dimensional turbulence and rigourously quantify it with physics based tests~\cite{king2018deep}, which introduces challenges both in dimensionality reduction and computational cost. Our initial results show considerable promise that these goals may be achievable and further efforts in this direction can be fruitful. We now present the datasets that would be used to test our approach, followed by the details and results in subsequent sections.

\subsection{Datasets}
\label{datadescription}
In order to assess the capability of our deep learning approach in emulating the spatio-temporal dynamics of turbulence, we choose two three-dimensional, unsteady flow datasets which have been studied well in literature, making them ideal subjects for this study. These datasets are briefly outlined below.

\subsection{Forced Isotropic Turbulence}
This dataset consists of a 3D Direct Numerical Simulation (DNS) of homogeneous, isotropic turbulence (HIT) in a box of size $128^{3}$. The flow is incompressible and statistically stationary in time. The flow is evolved in the simulation by constant energy forcing at low wavenumbers (corresponding to large scales), so that the flow does not decay with time. The data is generated from the \textit{spectralDNS}~\cite{mortensen2016massively} code which consists of a spectral solver with Fourier-Galerkin discretization.

\subsection{Forced Isotropic Turbulence with Passive Scalars}
This dataset is also HIT as the previous dataset, with the notable addition of two passive scalars in the flow. Furthermore, the numerical method used to generate this dataset is different, since the passive scalar forcing with pre-defined properties has to be accounted for. While the HIT is computed using a spectral solver with low-wavenumber forcing as before, the scalars are modeled using the reaction analogy (RA) method of Daniel et. al~\cite{daniel2018reaction}. This method mathematically models chemical reactions occurring between hypothetical reactants identified in a mixed fluid state. The reaction form ensures that the forcing term satisfies mass conservation, is smooth in the scalar space, enabling us to strictly enforce preset bounds and statistics for the two scalar fields.
The reaction analogy is useful to study flow-fields in nonpremixed combustion (the mixing process can be studied through the scalars) and non-Gaussian scalar turbulence. Therefore, it is included as a unique test case since the RA method mimics turbulent mixing and reacting flow in a well known HIT base flow, which we attempt to model with our deep learning approach.  We henceforth refer to this dataset as ScalarHIT. Both the HIT and ScalarHIT datasets together provide candidate testbeds in modeling both advection, transport and reacting flow.


\subsection{Statistical Diagnostics for Turbulence}
\label{diagnostics}
The accuracy of  machine learning predicted spatial fields is often assessed via standard statistical metrics such as mean square error and higher order statistical moments. However, it is critical to employ accuracy metrics which  reliably depict the \textit{physics} of the system generating the data. Such physics-based statistical metrics can provide useful, realistic and intuitive insight into performance of the predictive methods. Furthermore, such physical ``diagnostics" can inform us where ML methods would succeed and fail in real-world datasets, thereby driving further improvement in neural network architectures.

The statistical diagnostics we perform are (a) spectrum of energy fluctuations over scales, (b) statistics of velocity gradient  and (d) statistics of the coarse-grained velocity-strain alignment represented in the plane of the coarse-grained velocity gradient tensor. Together, these diagnostics enable our comparitive analysis of the machine learning predicted flow with the actual flow. The details of the various tests can be found in Appendix.

\section{Existing Approach: Convolutional LSTM Neural Networks}
\label{sec:convlstmintro}
Sequence prediction is different from other types of learning problems, since it imposes an order on the observations that must be preserved when training models and making predictions. Recurrent Neural Networks (RNNs) are a class of neural networks specially designed for such problems, which preserve this sequential information in the function being learned. The Long Short-Term Memory (LSTM) neural network is a special variant of RNN, which overcomes stability bottlenecks encountered in traditional RNNs (like the Vanishing Gradient), enabling its practical application. LSTMs can also learn and harness sequential dependence from the data, such that the predictions are conditional on the recent context in the input sequence. For instance, to predict the realization at time $t_i$, LSTMs can learn from the data at $t_{i-1}$ \textit{and also at times $t_{i-k}$}, where $k$ can be any number signifying the length of the prior sequence. In effect, $k$ represents the ``memory" in the system i.e. the extent to which the outcome of the system depends on its previous realizations. The LSTM networks are specialized at capturing these memory effects.

\begin{figure}
	\centering
	\fbox{\includegraphics[width=0.65\linewidth]{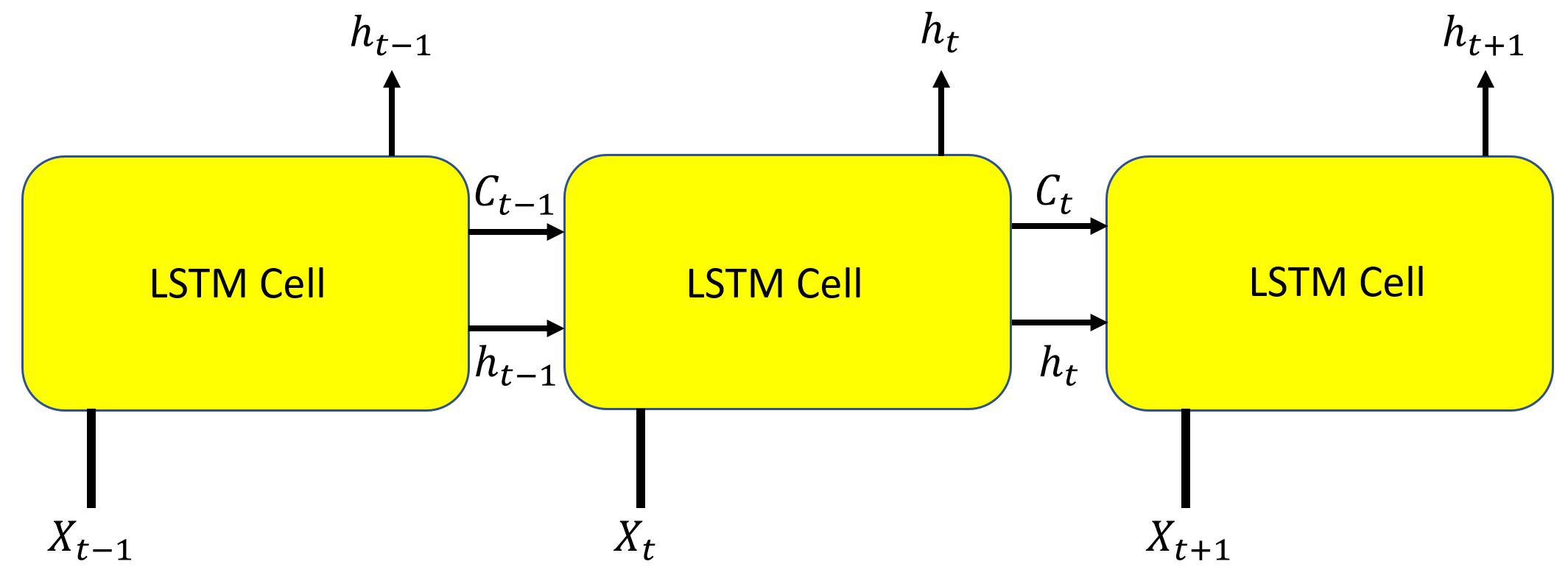}}
	\caption{LSTM Layout with Cell Connections}
	\label{lstmchain}
\end{figure}

The basic architecture of the LSTM NN is now outlined. The LSTM networks are different from other deep learning architectures like convolutional neural networks (CNNs), in that the typical LSTM cell contains three \textit{gates}: The \textbf{input} gate, \textbf{output} gate and the \textbf{forget} gate. The LSTM regulates the flow of training information through these gates by selectively adding information (input gate), removing (forget gate) or letting it through to the next cell (output gate). A schematic of the cells connected in a recurrent form is shown in Fig.~\ref{lstmchain}.

\begin{figure}
	\centering
	\fbox{\includegraphics[width=0.65\linewidth]{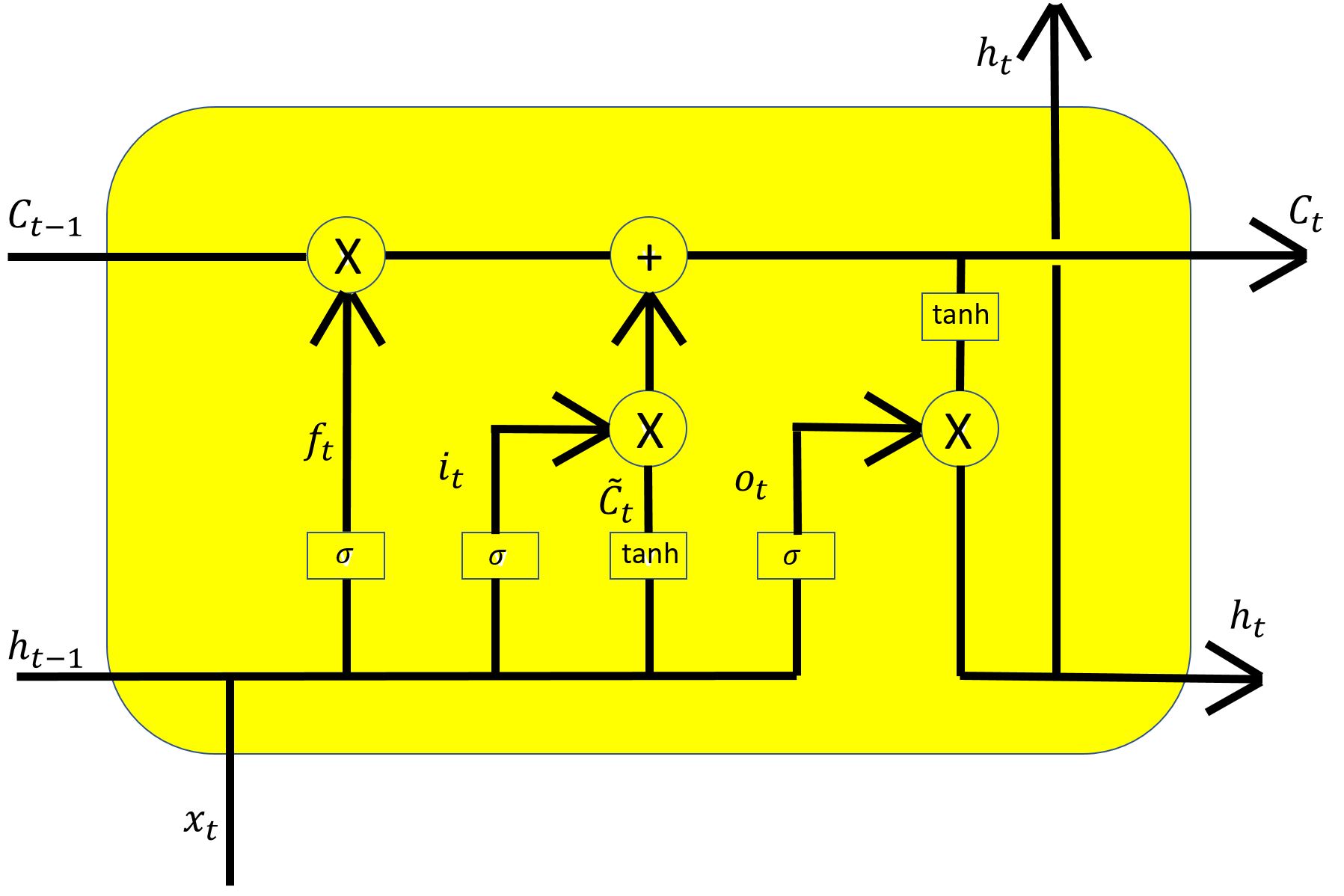}}
	\caption{Architecture of a LSTM Cell with Various Gates}
	\label{lstmcell}
\end{figure}

The input gate is represented by $i$, output gate by $o$ and forget gate by $f$. The cell state is represented as $C$ and the cell output is given by $h$, while the cell input is denoted as $x$. Consider the equations of a LSTM cell to compute its gates and states in Eqn~\ref{lstmeqn} and a schematic of its structure in Fig.~\ref{lstmcell}.

\begin{align}
    f_{t}\,&=\,\sigma \left(W_{f} \cdot \left[h_{t-1},x_{t}\right] + b_{f} \right) \nonumber \\
    i_{t}\,&=\,\sigma \left(W_{i} \cdot  \left[h_{t-1},x_{t}\right] + b_{i} \right) \nonumber\\
    \Tilde{C}_{t}\,&=\, tanh\left(W_{C} \cdot  \left[h_{t-1},x_{t}\right] + b_{C} \right) \nonumber\\    
    C_{t} \,&=\, f_{t}*C_{t-1} + i_{t}*\Tilde{C}_{t} \nonumber\\
    o_{t}\,&=\,\sigma \left(W_{o} \cdot \left[h_{t-1},x_{t}\right] + b_{o} \right)  \nonumber\\    
    h_{t} \,&=\, o_{t}*tanh \left(C_{t}\right)
    \label{lstmeqn}
\end{align}

$W$ are the weights for each of the gates and $\Tilde{C}$ is the updated cell state. These states are propagated ahead through the network, as shown in Fig.~\ref{lstmchain} and weights are updated by backpropagation through time. The forget gate plays a crucial role in reducing over-fitting by not retaining all information from the previous time steps. This arrangement of gates and selective information control is also the key reason why LSTMs do not suffer from the vanishing gradient problem which plagued traditional RNNs~\cite{hochreiter1998vanishing}. As a result, LSTMs are a powerful tool to model sequential datasets. A more detailed and introduction to LSTM can be found in Ref.~\cite{hochreiter1997long}.

In a traditional LSTM, the input and hidden states consist of an one-dimensional vector, therefore a two-dimensional input (such as an image or a data field) has to be resized to a single dimension. The "removal" of this dimensionality information fails to capture spatial correlations that may exist in such data, leading to increased prediction errors. In contrast, the Convolutional LSTM (ConvLSTM) was proposed by X Shi~\cite{xingjian2015convolutional} to process hidden and input states in two dimensions, thereby retaining spatial information in the data. ConvLSTM consists of a simple but powerful idea - that the gates in Eqn. ~\ref{lstmeqn} \textit{have the same dimensionality of the input data}. This enables us to provide a 2D image input and obtain 2D vectors $C_{t}$ and $h_{t}$ as outputs from the ConvLSTM cell. With this abstraction, the same equations in Eqn.~\ref{lstmeqn} can be used for the ConvLSTM cell, with the only difference being that the input vector and the cell gates have the same dimensionality. ConvLSTM has been successfully demonstrated for several sequential image prediction/classification tasks~\cite{wu2017convolutional}. Inspired by its success, we extend the ConvLSTM architecture to three dimensions. This is crucial for several applications in computational physics, as will be seen in the upcoming sections.

\section{Proposed Approach: Compressed Convolutional LSTM 3D}

Several real-world applications in turbulence consist of three dimensional dynamics, which make them expensive to compute with CFD based methods. As explained in the previous section, the goal of this work is to learn the attractor of the 3D DNS turbulent flow-field with Convolutional LSTM (ConvLSTM), and use it to model subsequent realizations. Therefore ConvLSTM method needs to be extended to 3D. However, we find that extension to 3D - though mathematically straightforward - creates severe constraints. This is because the large number of degrees of freedom in 3D dramatically increases the number of training parameters due to unrolling of recurrent networks. This has a significant impact on both the training time and computing cost, especially with limited GPU memory.

\subsection{Network Architecture}

To alleviate these difficulties, we propose a training methodology \textit{Compressed ConvLSTM3D} (CC-LSTM) - where we reduce the size of the training dataset while simultaneously retaining the same number of dimensions as the original data. We accomplish this by first ``compressing” the dataset to its \textit{latent space} by training a Convolutional Autoencoder (CAE). The encoder learns generalized weights to compress every snapshot of the dataset, while the decoder learns weights to ``decompress” the snapshot back into its original size. Therefore, we can choose the size of the encoded snapshot to be much smaller than the original snapshot, to attain the desired level of compression. We can use the trained encoder network to dynamically compress the input data to the CC-LSTM network, which in turn produces spatio-temporal predictions of the compressed state. These states are then dynamically decompressed back to the original size using the CAE decoder. 

\begin{figure}[t]
\includegraphics[width=10cm]{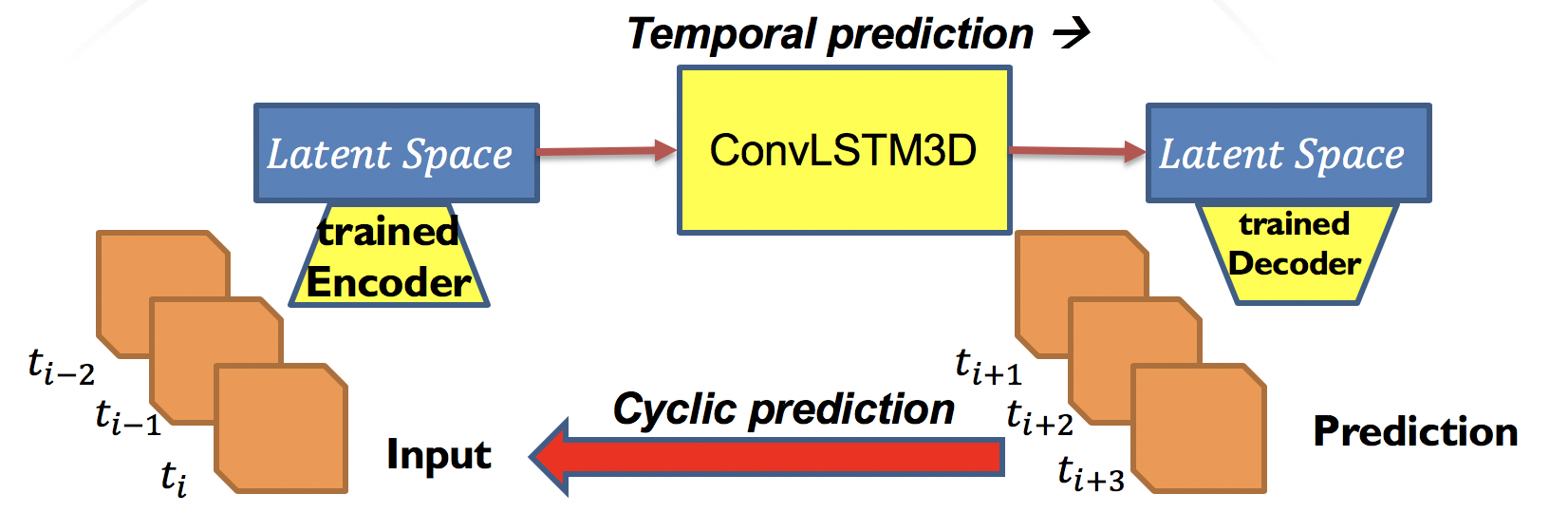}
\centering
\caption{Compressed ConvLSTM (CC-LSTM) Architecture}
\label{cclstm-schematic}
\end{figure}

A schematic of CC-LSTM network architecture is shown in Fig.~\ref{cclstm-schematic}. Furthermore, since LSTMs predict a fixed number of sequences (based on network design), continuous prediction can be obtained by ``seeding" the CC-LSTM with an initial input sequence, followed by using the network prediction as the next input sequence. The CC-LSTM produces long time predictions when this operation is performed in a cyclic manner. It should be noted that the poor predictive accuracy would quickly destabilize the long term predictions, and therefore this should typically be done when the network has trained with a sufficient level of accuracy. A major advantage of CC-LSTM is that it leads to massive reduction in the number of trainable parameters, while its latent space still retains key spatial correlations useful for reconstruction back to its original size. The information content, and therefore the quality of compression, in the latent space depends on the CAE architecture and training, and this will be explored in detail in the next section. 

\subsection{Training Strategy}
The fundamental idea underpinning data-driven modeling for complex systems is that there exists an attractor which governs the evolution of the dynamics with time. This attractor is typically low dimensional (for a high dimensional system), and explicitly knowing this attractor can enable us to make predictions at future states. A key hypothesis often used in dynamical systems theory is that the realizations of the system state contains information about this attractor, in form of its observables. Therefore, several studies have focused on estimating/approximating this attractor from observations, followed by reduced order models built from the computed attractors for system modeling. In turbulence, a popular strategy is to compute Proper Orthogonal Decomposition (POD) modes of the flow, which contain dominant dynamics of the flow in a smaller sub-space compared to that of the entire flow. These dominant modes are then evolved via Galerkin projection~\cite{noack2005need}, which projects the modal dynamics on the low dimensional subspace, thereby emulating the evolution of the flow's intrinsic low dimensional attractor. A more recent innovation has been to utilize Koopman operator theory to learn the low dimensional dynamics by directly learning the eigenpairs of the system~\cite{lusch2018deep,yeung2017learning}. However, Galerkin projection based approaches are known to have several limitations, chief among them being temporal stability. Recently, a deep learning approach was demonstrated by Mohan~\cite{mohan2018deep} where the POD modes were evolved with LSTM neural networks instead of Galerkin projection for a case of full turbulent homogeneous, isotropic flow. The results showed promise the ability of LSTM networks to capture non-linear, non-stationary dynamics in temporal evolution. The deep learning architecture proposed in the present work significantly extends this capability to include 3D spatio-temporal dynamics, in a compute efficient manner.

In case of the turbulence datasets described described in section~\ref{datadescription}, 
training is accomplished by a two-step process. The first step is reducing the dimensionality of the flow, since using the full flow information can be computationally prohibitive. Additionally, a justification can be made that the fundamental dynamics of the attractor are likely low dimensional, and hence extracting this information is necessary to learn the compact temporal evolution of the attractor; much like the use of POD and Koopman modes in literature. To achieve this dimensionality reduction, we employ the convolutional autoencoder (CAE), which learns compressed, low dimensional `` \textit{latent space}" representations for each snapshot in the flow. The CAE has two main components - the \textit{encoder} and the \textit{decoder}. The representational information to transform the snapshot from its original high-dimensional state to the latent space is stored in the encoder. Similarly, the inverse transformation from the latent to original state is learned by the decoder. Both the encoder and decoder are matrices which are learned by standard neural network backpropagation and optimization techniques. It is important to note that this is a \textit{convolutional} autoencoder, such that the spatial information is learned by moving filters, as in a convolutional neural network. The moving filters capture various spatial correlations and drastically reduce the number of weights we need to learn (and thereby the sizes of the encoder, decoder matrices) due to parameter-sharing. This makes the training considerably cost effective and faster than using a standard fully-connected autoencoder. The reader is referred to Ref.~\cite{goodfellow2016deep} for more details.

The second step is to learn the spatio-temporal evolution of the low dimensional latent space. This is accomplished using the Convolutional LSTM cells shown shown before. As before, we use Convolutional layers as input to the LSTM cells since they can utilize the spatial correlations directly without resorting to the expensive fully connected architecture. This enables us to use most of the training effort in learning the temporal dynamics governing the latent spaces. We reiterate here that the using the ConvLSTM cell is memory intensive and the latent space representation makes the training tractable to be handled on a single GPU, making it a highly efficient learning framework.

\section{Dimensionality Reduction with Convolutional Autoencoder}
A key advantage of the CC-LSTM architecture is the dimensionality reduction of the large dataset to be modeled. This enables us to develop a ConvLSTM model for such datasets at a very low computational cost. However, dimensionality reduction can be a lossy compression process - for instance, popular approaches like Proper Orthogonal Decomposition can represent dominant energetic dynamics in the first few eigenpairs. These eigenpairs can be used for further analysis or modeling tasks, such as Galerkin projection, while the neglected eigenpairs have very low energy contribution to the overall dataset. The POD compresses the dataset as a \textit{linear} map, whereas autoencoders with multiple layers and non-linear activation functions  are inherently \textit{non-linear} maps. As a result, autoencoders can provide very high compression ratios (defined in the next section) for the same dataset. The architecture of the CAE and its physical connections are outlined below.

\subsection{CAE Architecture}
\label{CAEdetail}
\begin{figure}[t]
\includegraphics[width=12cm]{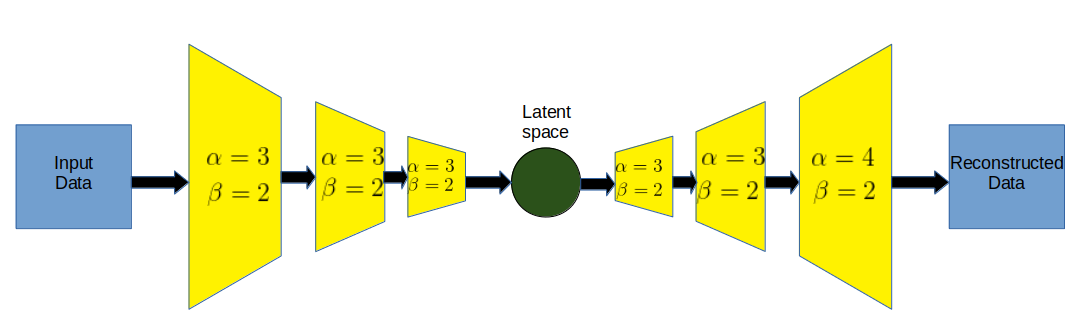}
\centering
\caption{Convolutional Autoencoder Architecture for HIT}
\label{cae-hit-schematic}
\end{figure}

\begin{figure}[t]
\includegraphics[width=12cm]{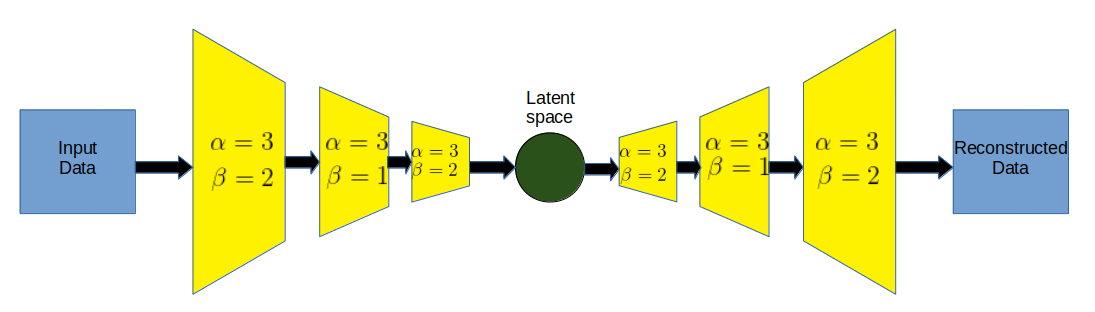}
\centering
\caption{Convolutional Autoencoder Architecture for ScalarHIT}
\label{cae-scalarhit-schematic}
\end{figure}

Schematics of the CAE architecture used for the HIT and ScalarHIT datasets are shown in Figs.~\ref{cae-hit-schematic} and \ref{cae-scalarhit-schematic} respectively. The convolutional autoencoder is different from a standard autoencoder, where the fully connected weights are replaced by shared weights in a convolutional kernel. This greatly reduces the memory utilization in such networks, since the same $n$ weights in a convolutional kernel are translated throughout the domain of size $m \times m$, where $m >> n$. These $n$ weights are global, hence learned for all regions of the domain. In contrast, the standard autoencoder architecture would need $m^{2}$ weights which are local, leading to prohibitive memory consumption and extremely slow training. In addition to computational benefits, the design of the convolutional kernel offers flexibility in tuning the number of shared weights and mode of translation through the domain. Two key choices have to be made: the \textit{kernel size}, $\alpha$ and \textit{kernel stride}, $\beta$. The kernel size indicates the spatial extent of a single kernel. For instance, a choice of $3 \times 3$ creates 9 shared, trainable weights in the kernel. The next choice is to decide how the shared weights (i.e. the kernel) translate across the domain. An illustration is shown in Fig.~\ref{striding}, where a kernel $\alpha=3$ can be translated by a distance of our choosing, known as stride. The figure shows kernel positions after 1 or 2 strides on the domain, and the strides are repeated until the entire domain has been exposed to the kernel.

At the core of a CAE is the convolution operation, which already has mature applications in fluid mechanics such as subgrid model development for Large Eddy Simulations. In the CAE encoder network (i.e. layers to the left of the latent space in the schematics), the kernel convolves with the data to reduce its dimensionality for every time instant $t_{i}$. As a result, a $\alpha = 3$ kernel downsamples a spatial field of the same size- known as the \textit{receptive field} (Ref.~\cite{goodfellow2016deep})-  to a single point. The decoder network kernel (i.e. layers to the right of the latent space in the schematics) then upsamples each point in the latent space back to the size of the receptive field through a deconvolution operation. Downsampling in the case of neural networks can be explained as a weighted averaging operation, where the averaging weights are learned. Similarly, the upsampling kernel weights are also learned to perform the inverse operation. By stacking multiple layers of encoders, the input is downsampled in every layer and the resulting domain - the latent space - can be extremely low dimensional. Likewise, upsampling can be performed by suitable number of decoding layers to recover the original dimension.  For a fixed kernel size $\alpha$, the striding of the convolutional kernel has a direct effect on the dimensionality of the convolved output after each layer. From Fig.~\ref{striding}, it is clear that increasing the stride diminishes the coverage of kernel over the domain, making the convolved output sparser. For a fixed $\alpha=3$, $\beta = 1$ leads to an overlap with receptive field at the previous stride, while $\beta = 2$ removes any overlap. Higher values of $\beta$ create gaps in the domain which are not processed by the kernel, and hence can traverse the entire domain in fewer steps than using $\beta=1$. The impact of $\alpha$ and $\beta$ are discussed in the next section.

Another important aspect is the number of features in the input to CAE. In the case of HIT dataset, there are $3$ features corresponding to the $3$ components of velocity. Similarly the ScalarHIT dataset would have $5$ input features due to the addition of $2$ passive scalars. While the number of features should be the same at the encoder input and decoder output, it is not necessarily the case for the latent space. Increasing the number of features in the latent space accommodates more information at a minimal increase in computing cost. The compression ratio $z$ is thus defined as 

\begin{equation}
    z \,=\, \frac{(\mathrm{original \ dimensions \times number \ of \ input \ features})}{(\mathrm{latent \ dimensions \times number \ of \ latent \ features})}
    \label{compressionratio}
\end{equation}

From Equation~\ref{compressionratio}, it follows that for input dimensions of size $128^3$ with $3$ features; and latent space dimensions of $15^3$ with $15$ features, there is considerably lesser impact on $z$ with increase in latent space features. As such, the most significant effect comes from the latent space dimensions, giving us the liberty to increase the features. In fact, results with increased number of features had a direct effect on accuracy compared to having the same number of features as the input. In other words, the subspace spanned by the input features is mapped into a latent subspace spanned by a greater number of features.

\subsection{Physical Interpretation}
\begin{figure}
\centering
\includegraphics[width=6cm]{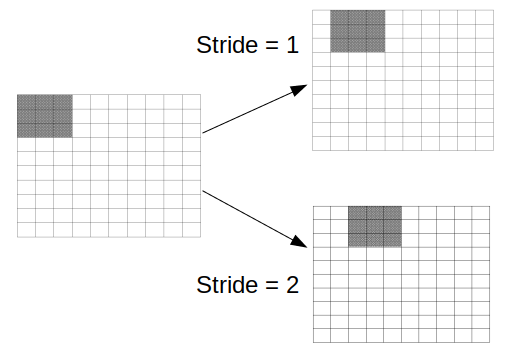}
\caption{Variable Striding in Convolutional Kernels}
\label{striding}
\end{figure}

It is now worthwhile to discuss implications of the these choices in dimensionality reduction of complex, spatio-temporal and multiscale datasets like turbulence. From the discussion above, it is apparent that there are two competing strategies for dimensionality reduction in a CAE. The first strategy relies on varying $\alpha$ to increase the receptive field. A larger receptive field would decompose several adjacent data-points into a single data-point. Therefore, for a desired dimension of the latent space, a suitable value of $\alpha$ can be computed. The second strategy is to retain a constant, low $\alpha$, but increase $\beta$ to traverse the domain in as few steps as possible. The optimum $\beta$ can be estimated from the desired latent space dimension. There are caveats to both these strategies: A larger receptive field; in the limit of $\alpha \rightarrow \infty$ increases the number of trainable weights, with their number approaching the number of data-points in the domain. As mentioned before, this is computationally prohibitive for 3D datasets of even small sizes and is hence not feasible. Therefore, a low $\alpha$ would be practical. This leads to the second strategy of increasing $\beta$, which also has significant pitfalls due to large discontinuities created between adjacent receptive fields. A low $\beta$ leads to minimal discontinuities in convolution operations between adjacent receptive fields. In contrast, numerical artifacts increase with higher $\beta$ due to sudden discontinuities/gradients between the receptive fields. Additionally, higher $\beta$ also skips over some features in the domain, causing loss of accuracy in the latent space. This is illustrated in Fig.~\ref{striding}, where $\beta=1$ for $\alpha=3$ ensures maximum overlap, and hence the weights can better capture the spatial correlations. In essence, there is an upper limit on $\beta$ for a given $\alpha$, beyond which significant numerical errors may manifest.

\begin{figure}
\includegraphics[width=2cm]{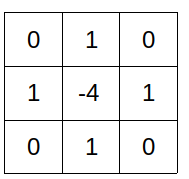}
\centering
\caption{Convolutional Kernel for $2^{nd}$ Order Central Differencing Numerical Stencil}
\label{stencil}
\end{figure}

At this juncture, it is useful to develop some intuition on $\alpha$ and $\beta$ in terms of numerical solution of fluids. The convolutional kernel used in CAE has direct connections to the numerical stencils used in finite difference/finite volume approaches. Consider the standard $2^{nd}$ order central difference scheme in one dimension for a quantity $\phi$
\begin{equation}
    \frac{\phi_{i+1} - 2 \phi_{i} + \phi_{i+1}}{\delta^{2}}
\end{equation}
In two dimensions, this can be represented as convolutional kernel of $\alpha=3$ with constant weights as seen Fig.~\ref{stencil}. In the CAE, all the constant weights in the convolutional kernels are replaced with learnable weights. As such, the output of trained kernel can be thought of as a weighted combination of adjacent points, akin to numerical solution of PDEs. In fact there, are deeper connections between convolutional kernels and  stencils of numerical schemes that have been uncovered recently for developing efficient neural network based PDE solvers, and the reader is directed to the works of Dong~\cite{long2017pde,dong2017image}.
In numerical solutions of PDEs, the kernel corresponds to the order of numerical scheme, which is typically constant and computed at \textit{every point in the domain}.
Thus, each layer of the CAE encoder consists of a customized numerical stencil specific to the dataset. By analogy, larger stencils may represent higher order numerical schemes, as seen by an increased number of trainable weights in networks. Seeking to make connections with Convolutional neural networks, a PDE solver has a constant $\alpha$ and $\beta=1$ in a single layer of operation. In contrast, the CAE has multiple layers of operation with flexibility to have different $\alpha,\beta$ in each layer. In lieu of these close connections, the major differences between PDE solvers and CAEs boil down to the treatment of boundaries, stride and the mapping of input features into a different subspace. In CAE, only the first layer in the deep neural network encoder treats the boundaries, while the increasing $\beta$ at successive layers dramatically cuts down on dimensionality. In summary, CAE encoders are able to encode high dimensions into low dimensions with an intelligent choice of kernel weights, kernel sizes and stride lengths. The CAE decoder is essentially an inverse operation of the encoder, but not in an implicit fashion~\cite{ardizzone2018analyzing}. Instead the decoder weights and strides are learned explicitly to compute the inverse map from latent space to original data. There are other implementation details in CAEs due to padding, which can also slightly affect $\alpha$ and $\beta$. 
The reader is referred to standard texts on deep learning for further details. 

Consider the CAE architectures for the HIT and ScalarHIT datasets in Figs~\ref{cae-hit-schematic} and \ref{cae-scalarhit-schematic} respectively. Based on the discussion above and Fig.~\ref{striding}, it follows that long strides can prevent the kernel from capturing flow features occupying small length scales and extremely longer strides can ignore even larger features. In terms of Kolmogorov spectra of turbulence, a strong argument can be made that choosing $\beta=1$ would capture turbulent structures at all wavenumbers while a $\beta = 2$ would neglect high wavenumbers but excel at capturing low wavenumbers. A further increase in beta would lead to degradation in capturing all wavenumbers. Since there are multiple layers, any loss in features seen in the first layer would be amplified. However, a small stride at every layer has a minimal effect on the dimensionality reduction of the data. It is possible to stack numerous layers with $\beta=1$ to achieve the desired low dimensional latent space, while also capturing all wavenumbers. However, the computational costs rise rapidly due to memory and difficulty in simultaneously training a large number of kernels. This runs antithetical to the goal of a reduced order model, so we make choices of $\beta$ which neglect high wavenumbers and reproduce scales in the large scale and inertial range scales, which can be trained at a very low cost. 

\subsection{Compression Performance}
\begin{figure}
    \centering
    \begin{subfigure}[t]{0.4\textwidth}
        \includegraphics[width=\textwidth]{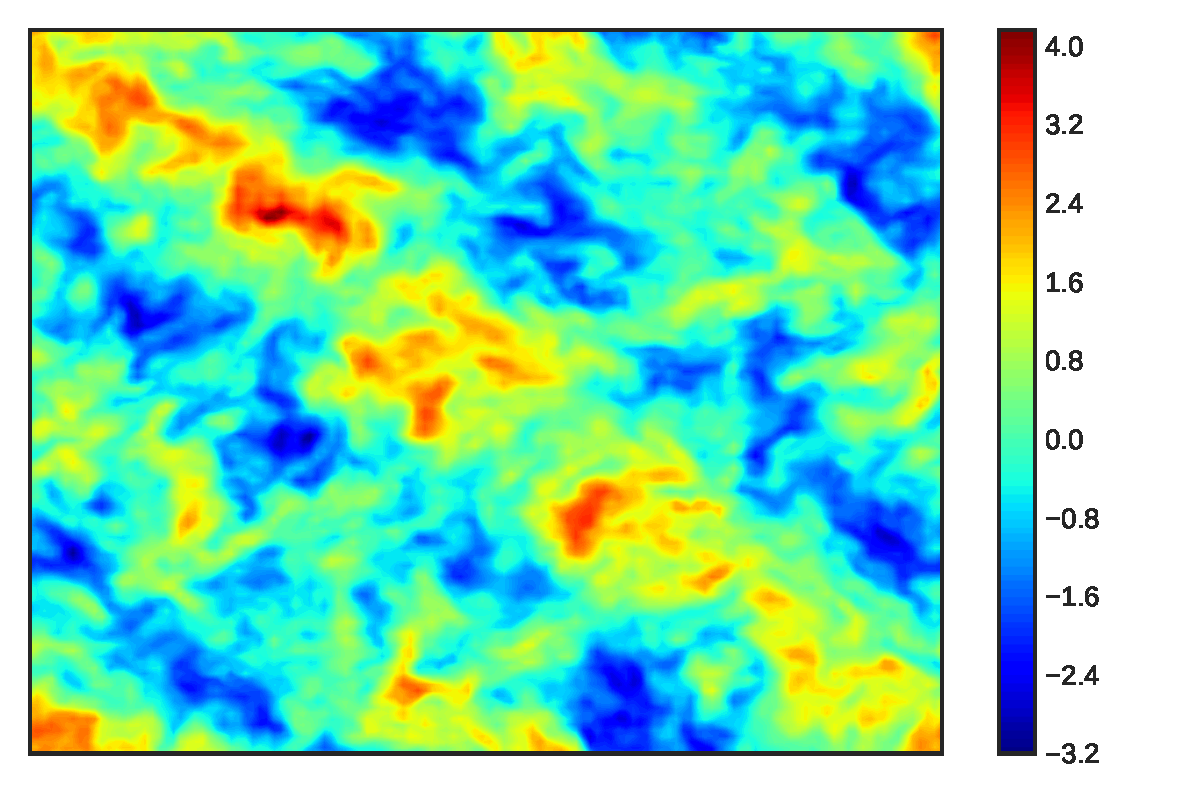}
        \caption{$U$ velocity}
        \label{CAEinput:U}
    \end{subfigure}
    ~ 
    \begin{subfigure}[t]{0.4\textwidth}
        \includegraphics[width=\textwidth]{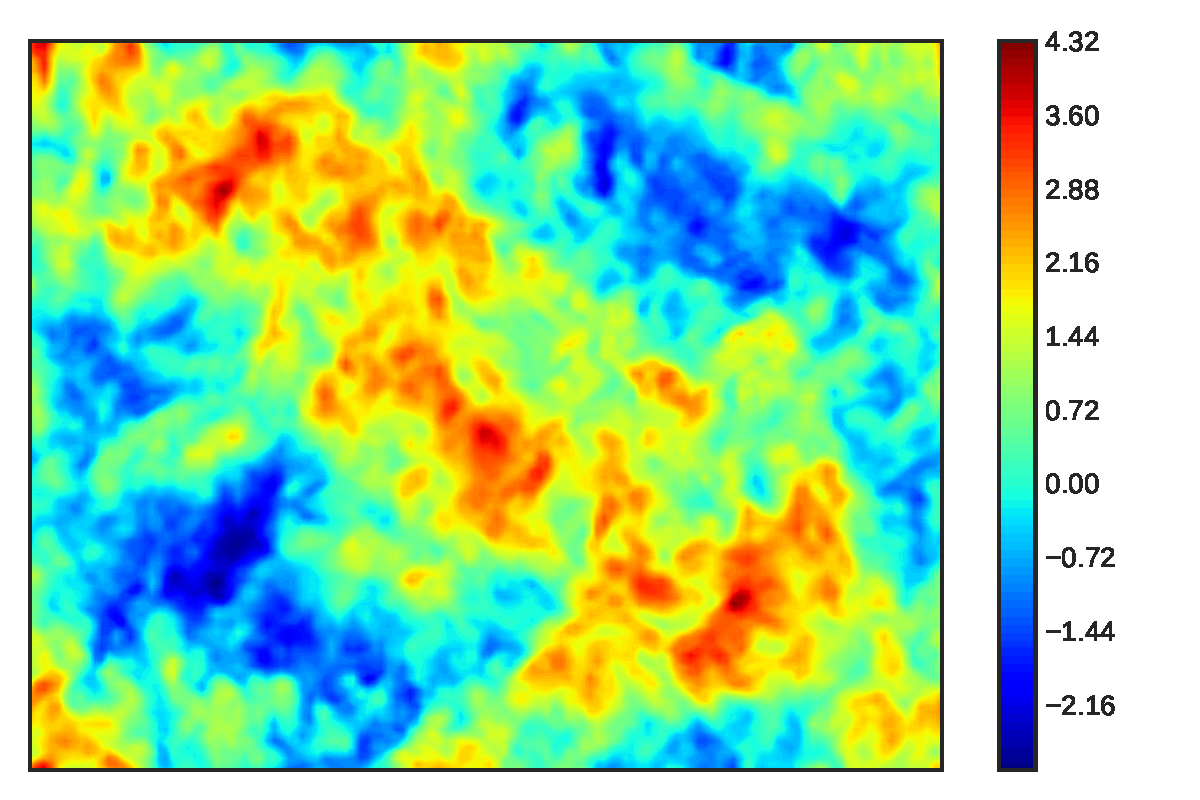}
        \caption{$V$ velocity}
        \label{CAEinput:V}
    \end{subfigure}
    ~ 
    \\
    \begin{subfigure}[t]{0.4\textwidth}
        \includegraphics[width=\textwidth]{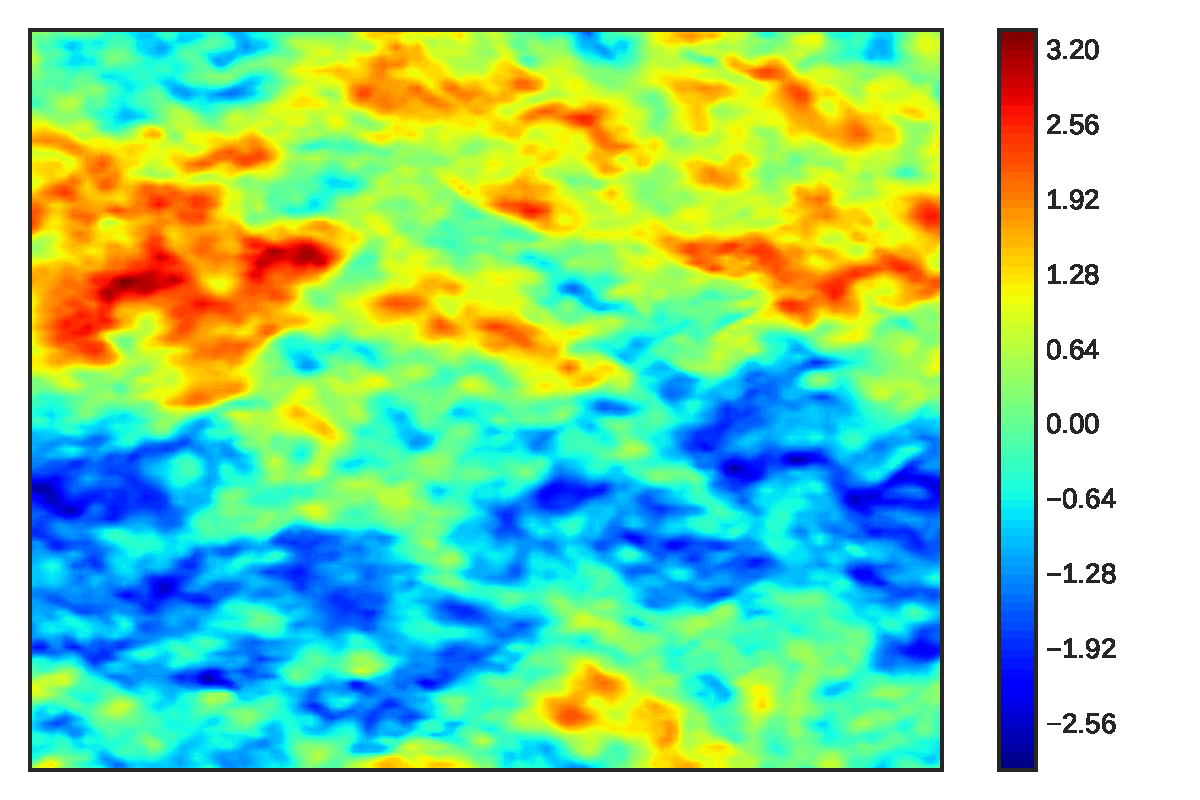}
        \caption{$W$ velocity}
        \label{CAEinput:W}
    \end{subfigure}
    \caption{$3\mathrm{D}$ HIT flow snapshot input to CAE with compression ratio $z=125$: $2\mathrm{D}$ slices}\label{CAEinputgraphics}
\end{figure}
\begin{figure}
    \centering
    \begin{subfigure}[t]{0.4\textwidth}
        \includegraphics[width=\textwidth]{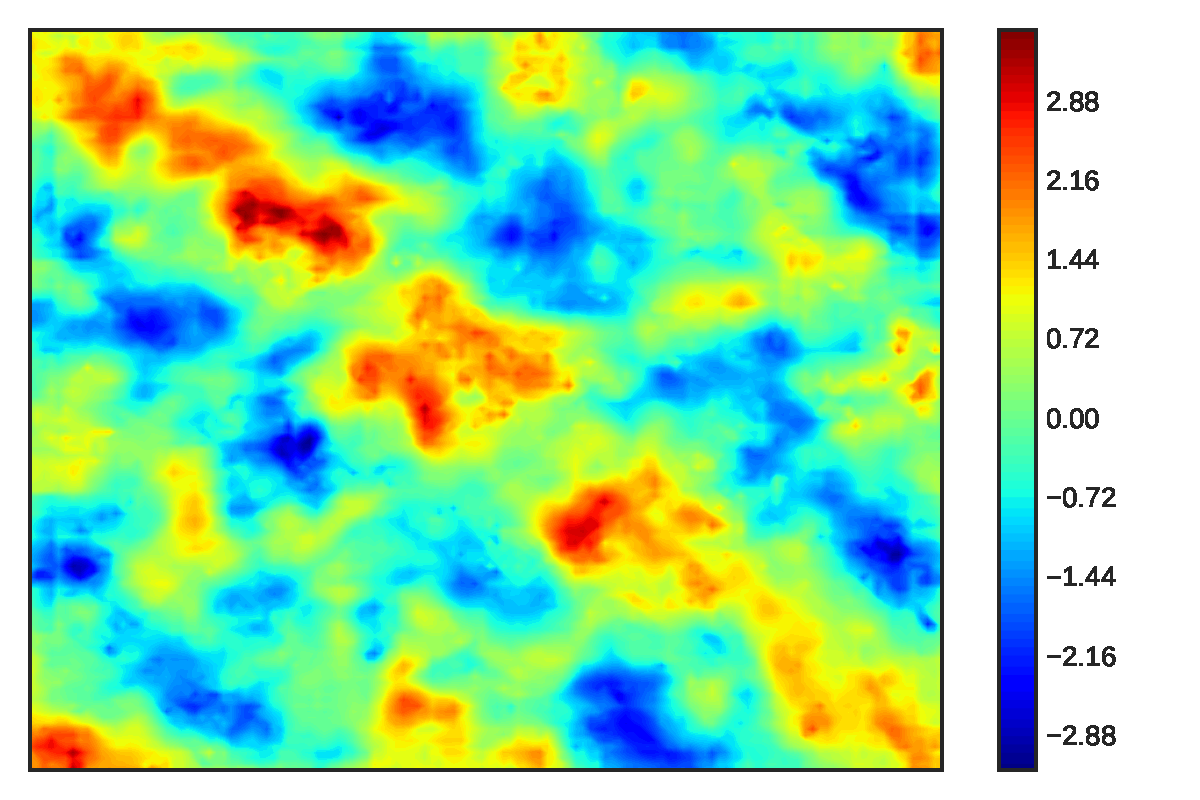}
        \caption{$U$ velocity}
        \label{CAEoutput:U}
    \end{subfigure}
    ~ 
    \begin{subfigure}[t]{0.4\textwidth}
        \includegraphics[width=\textwidth]{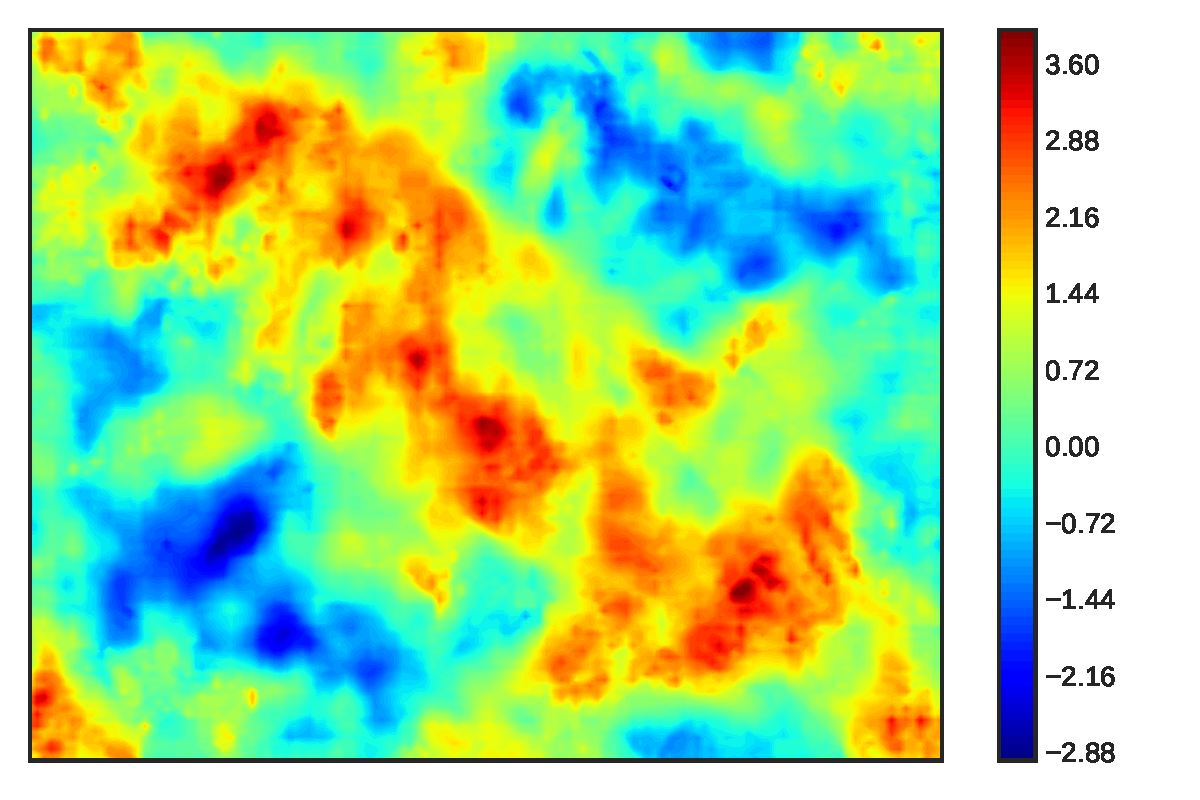}
        \caption{$V$ velocity}
        \label{CAEoutput:V}
    \end{subfigure}
    ~ 
    \\
    \begin{subfigure}[t]{0.4\textwidth}
        \includegraphics[width=\textwidth]{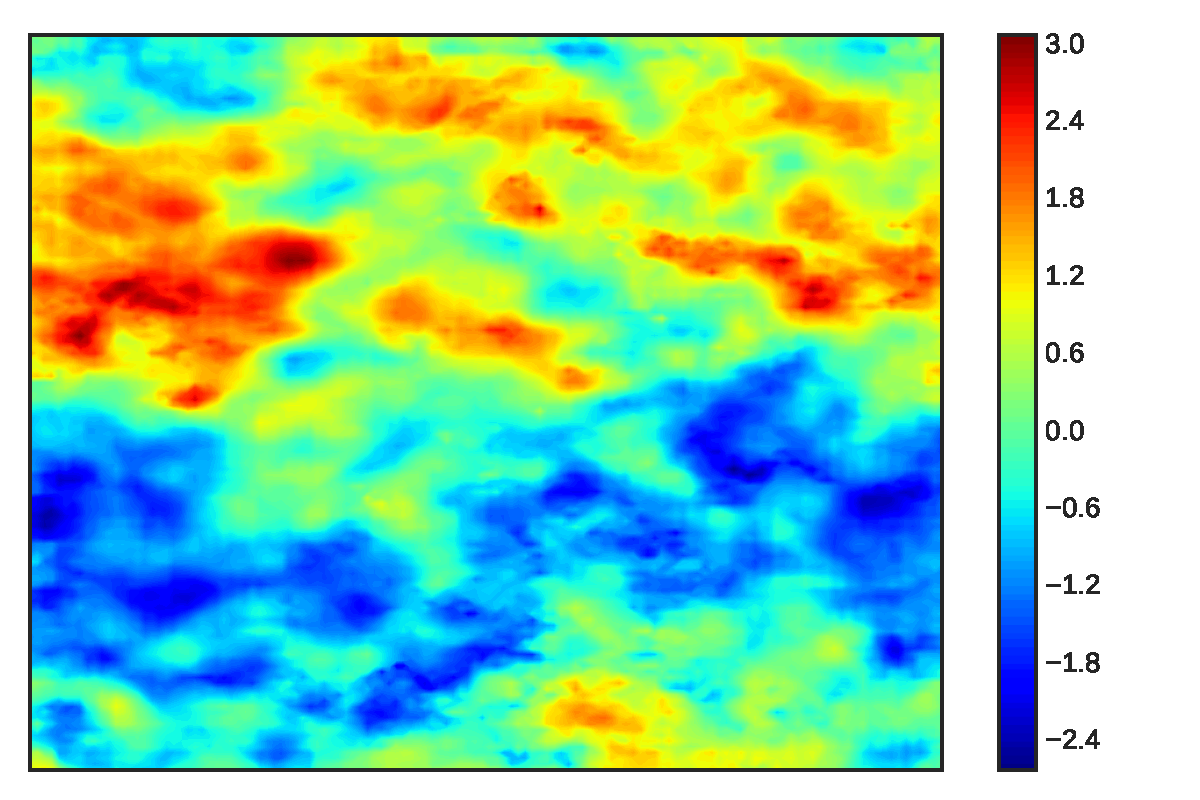}
        \caption{$W$ velocity}
        \label{CAEoutput:W}
    \end{subfigure}
    \caption{Compressed $3\mathrm{D}$ HIT flow snapshot with CAE compression ratio $z=125$: $2\mathrm{D}$ slices}\label{CAEoutputgraphics}
\end{figure}
\begin{figure}
    \centering
    \begin{subfigure}[t]{0.2\textwidth}
        \includegraphics[width=\textwidth]{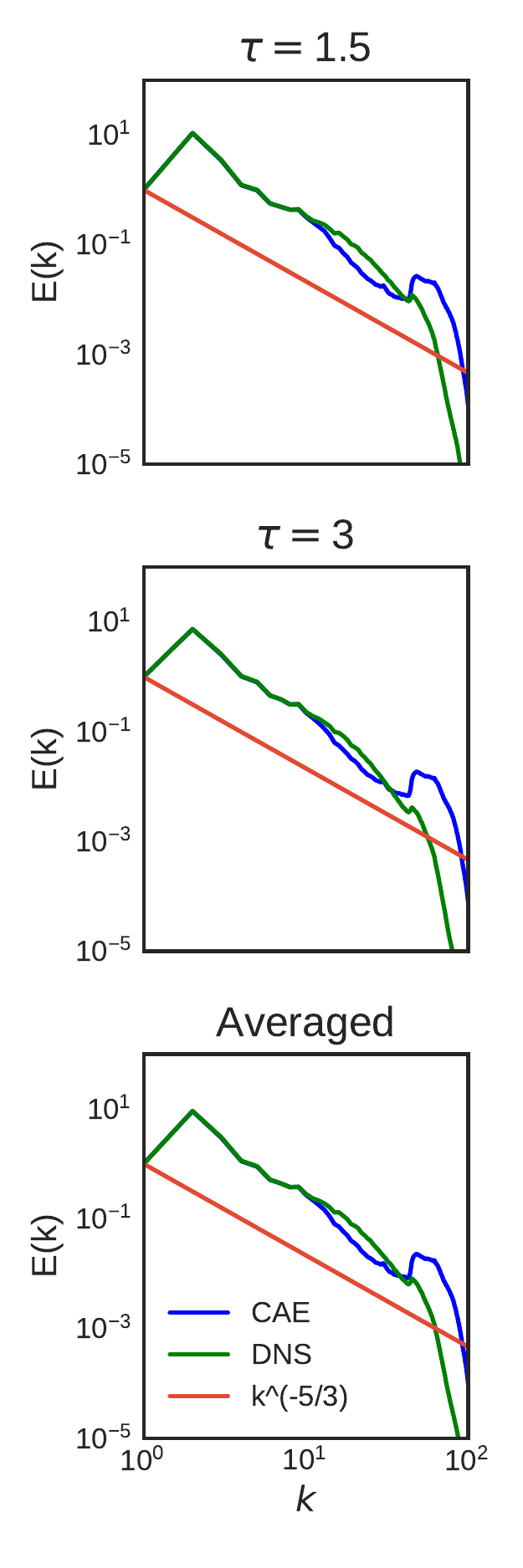}
        \caption{Energy Spectra}
        \label{HIT_Comp:spec}
    \end{subfigure}
    ~ 
    \begin{subfigure}[t]{0.2\textwidth}
        \includegraphics[width=\textwidth]{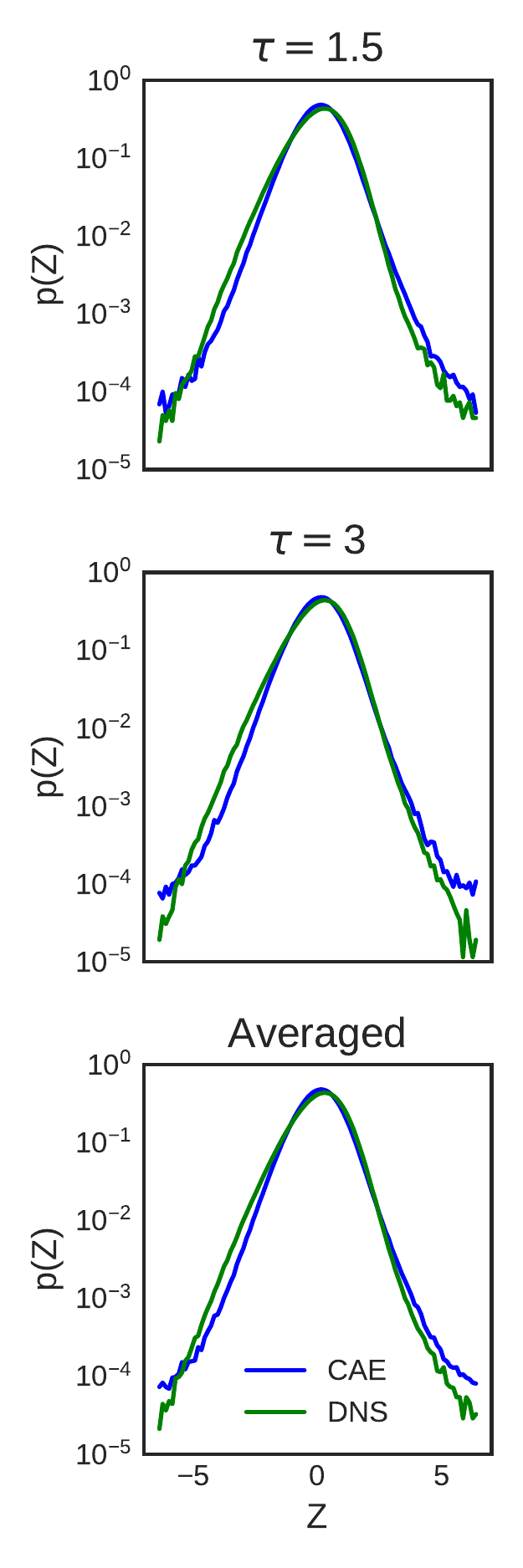}
        \caption{PDF of Velocity Gradients}
        \label{HIT_Comp:PDF}
    \end{subfigure}
    ~ 
    \\
    \begin{subfigure}[t]{0.4\textwidth}
        \includegraphics[width=\textwidth]{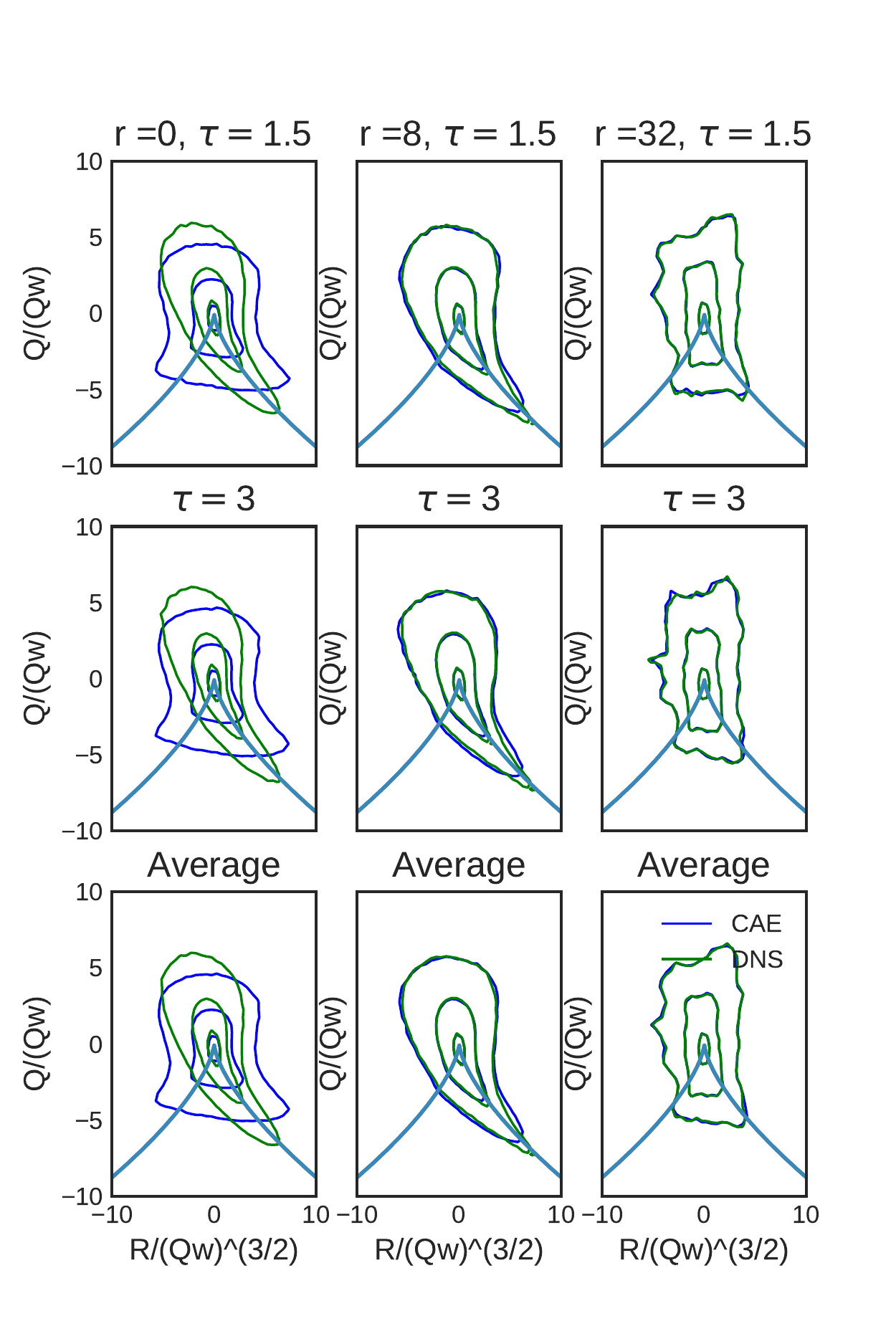}
        \caption{$Q-R$ plane at various coarse-grained scales}
        \label{HIT_Comp:qr}
    \end{subfigure}
    \caption{Statistics of Compressed HIT flow vs DNS - Compression $z\,=\,125$}\label{HIT_Comp}
\end{figure}
\begin{figure}
    \centering
    \begin{subfigure}[t]{0.2\textwidth}
        \includegraphics[width=\textwidth]{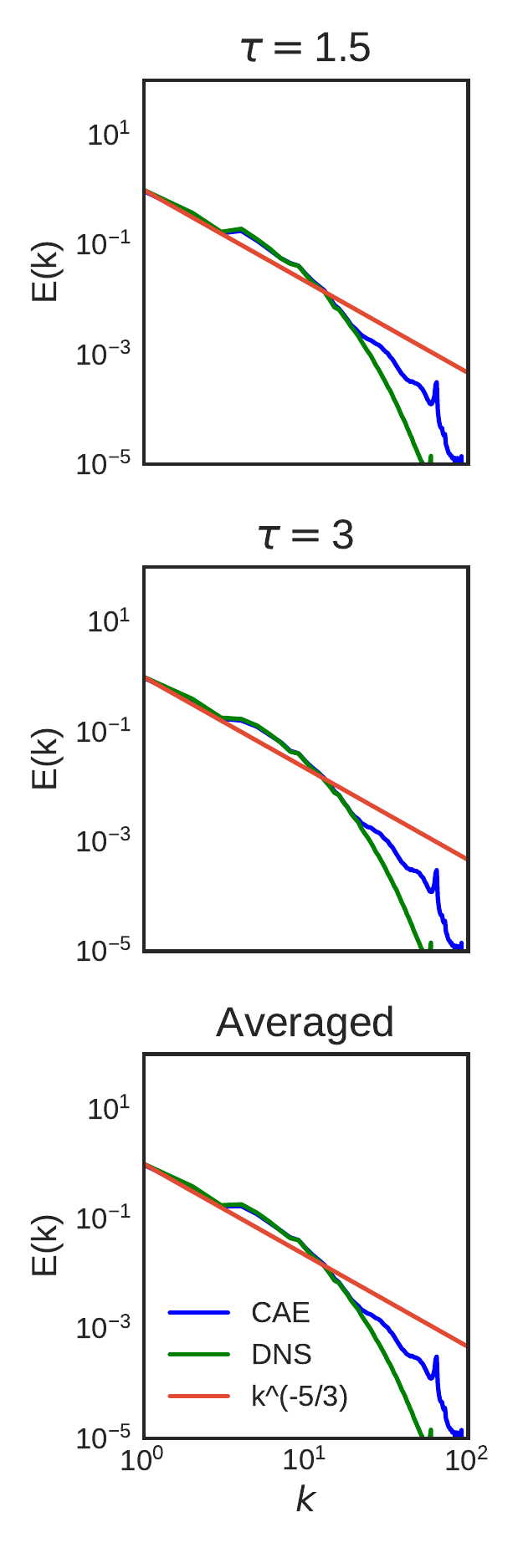}
        \caption{Energy Spectra}
        \label{ScalarHIT_Comp:spec}
    \end{subfigure}
    ~ 
    \begin{subfigure}[t]{0.2\textwidth}
        \includegraphics[width=\textwidth]{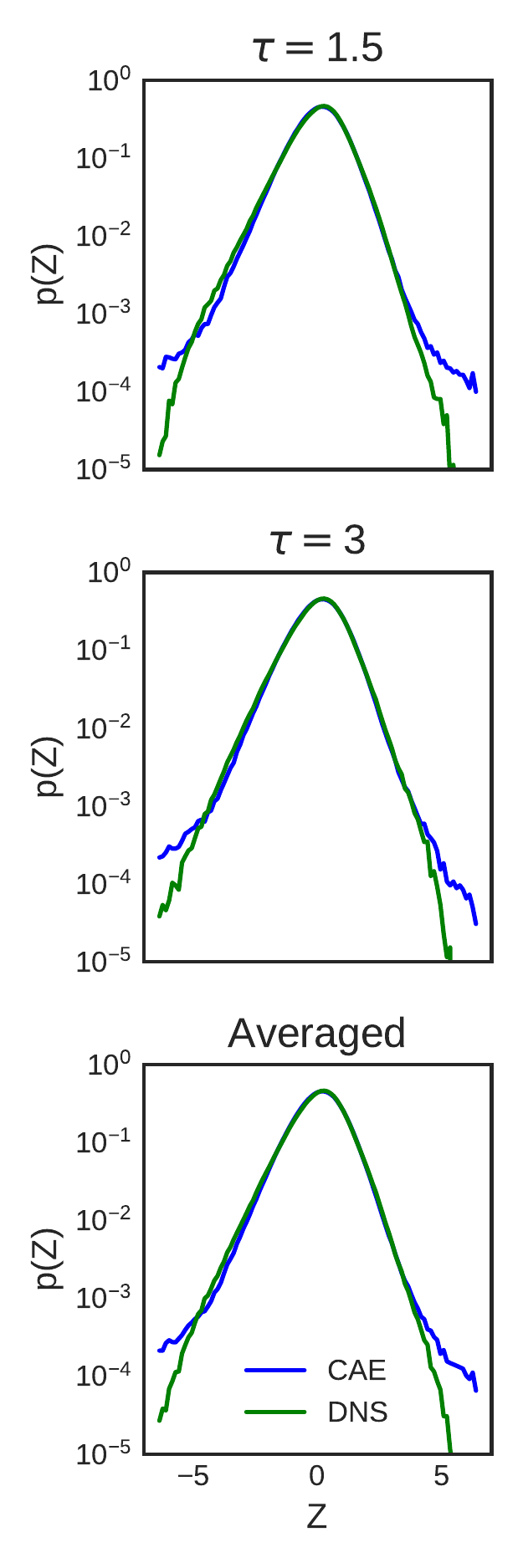}
        \caption{PDF of Velocity Gradients}
        \label{ScalarHIT_Comp:PDF}
    \end{subfigure}
    ~ 
    \\
    \begin{subfigure}[t]{0.4\textwidth}
        \includegraphics[width=\textwidth]{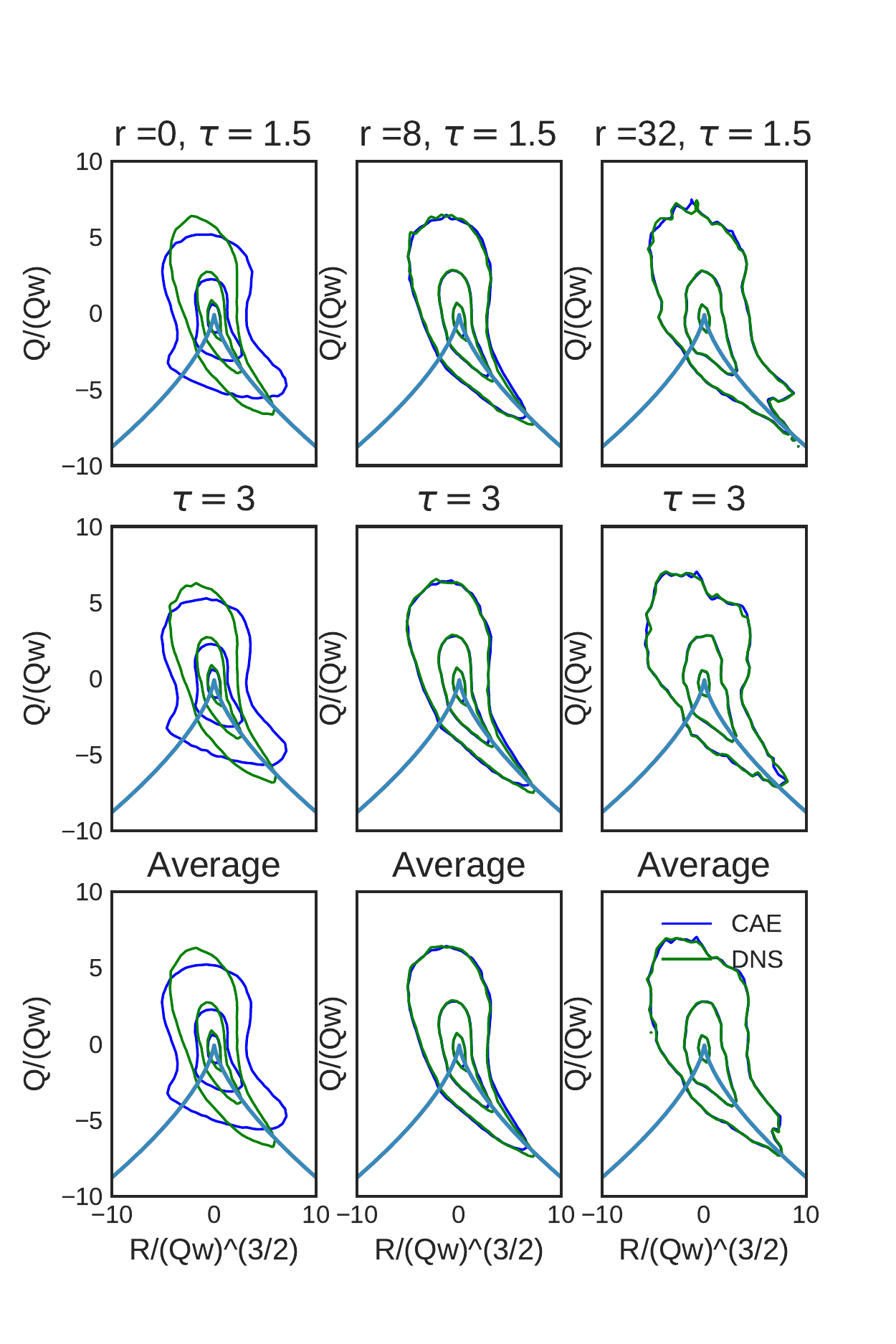}
        \caption{$Q-R$ plane at various coarse-grained scales}
        \label{ScalarHIT_Comp:qr}
    \end{subfigure}
    \caption{Statistics of Compressed ScalarHIT flow vs DNS - Compression $z\,=\,20$}\label{ScalarHIT_Comp}
\end{figure}

We now analyze the accuracy of the compressed dataset for both HIT and ScalarHIT. In order to highlight the effect of compression ratio on the reconstruction accuracy of the compressed dataset, the HIT and ScalarHIT datasets are compressed at different ratios, $z=125$ and $z=20$ respectively. We seek to investigate if our physics-driven design of the CAE indeed neglects primarily the small scale features of turbulence. In order to do so, the latent space for each dataset is reconstructed to its original dimensions using its CAE decoder.
As an example, a sample of the compressed flow is shown in Fig.~\ref{CAEoutputgraphics}, for the actual input flow in Fig.~\ref{CAEinputgraphics}. The CAE compression shows a high quality of reconstruction, especially for a fairly high compression ratio of $z=125$. While minor differences exist, a consistent approach to quantify the efficiency of CAE compression is through the turbulence statistical tests described in Section~\ref{diagnostics} and the appendix. The results for HIT and ScalarHIT are shown in Fig.~\ref{HIT_Comp} and Fig.~\ref{ScalarHIT_Comp} respectively.

The results for HIT show that the small scales are indeed neglected as we intended in the previous section. However, there also appear to be some losses in the inertial scales. As expected, we have excellent accuracy in the large scales, as seen from all three tests. The results for ScalarHIT show some marked differences.  The small scales are also neglected here as expected, but there appears to be minimal losses in the inertial scales. This is apparent from the energy spectra in Fig.~\ref{ScalarHIT_Comp:spec} and also the near identical reconstruction for the inertial and large scale features in all three tests. In other words, the compression ratio has a non-trivial impact on the flow compression, but increases memory costs. This is a trade-off that must be made for data-driven modeling, and is typically dictated by the physics of interest in the desired model and the computational costs.

\section{CC-LSTM Results}
\label{results}
The primary objective of this work is to explore capability of deep neural networks to learn the underlying attractor of high dimensional systems, and use it to model subsequent realizations. In order to study this, we choose HIT and ScalarHIT datasets, which exhibit isotropic but highly multiscale, turbulent phenomena. The statistical stationarity of these datasets indicates that a suitable DL approach should capture the attractor dynamics from as few snapshots as possible. In this section, we discuss the accuracy of the learned attractor by analyzing the statistics of the predicted flow snapshots, and comment on its strengths and weaknesses.
The CC-LSTM architecture described in Sections~\ref{sec:convlstmintro} and \ref{CAEdetail} is are trained on $20$ sequential, 3D flow snapshots for both the datasets, corresponding to $1.25$ eddy turnover times. The network consists of $3$ Convolutional LSTM layers with $80$ hidden units in each layer. The final output layer has a linear activation unit with ReLU activation at the hidden layers. The data is scaled with zero mean and unit variance to improve learning.
A key user specified parameter in the ConvLSTM module is the time history used to make each set of predictions, as outlined in Section~\ref{sec:convlstmintro}. This parameter is referred to as $k$. A $k=5$ indicates the network uses only $5$ snapshots from the past to predict the next $5$ snapshots, and $k=1$ uses only the snapshot at the previous instant to predict the next snapshot i.e. Markovian-type dynamics. This choice can be made to reflect our intuition of the flow physics, and it is noted that a higher $k$ introduces more trainable parameters in the network, increasing computational cost. We use $k=3$ for both these datasets, and the validity of this choice will be discussed further in Section~\ref{k_variation}. In order to illustrate the effect of compression ratio on modeling accuracy, both the HIT and ScalarHIT are compressed with different ratios. 

\subsection{Homogeneous Isotropic Turbulence}
\begin{figure}
    \centering
    \begin{subfigure}[t]{0.4\textwidth}
        \includegraphics[width=\textwidth]{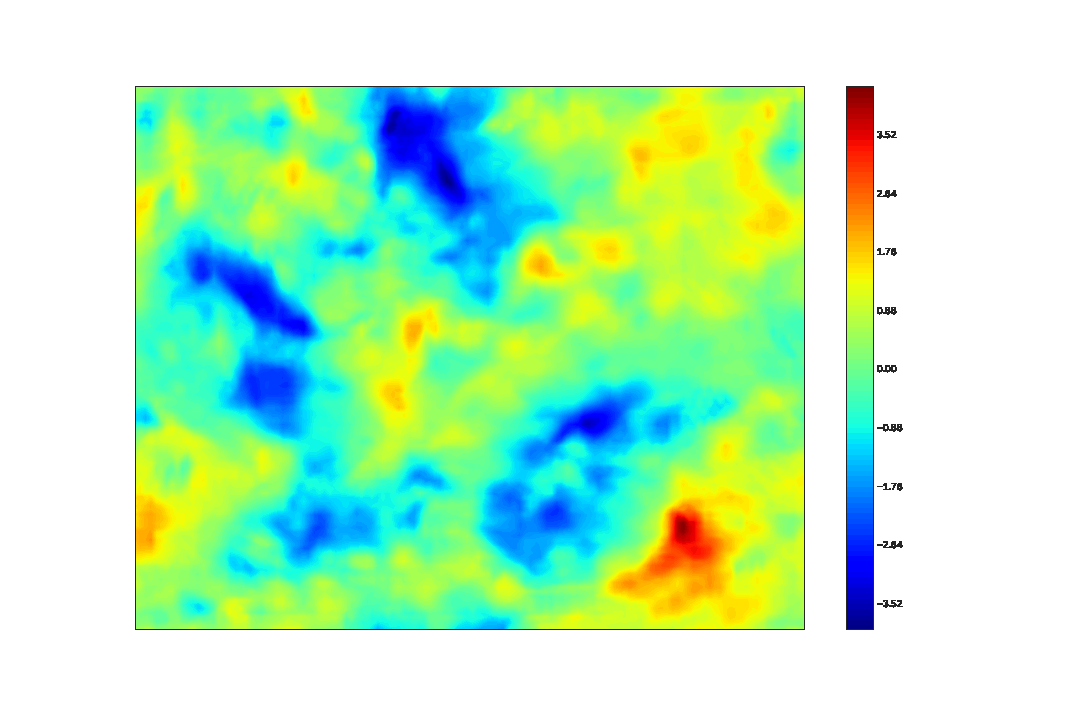}
        \caption{$U$ velocity}
        \label{modelgraphics:U}
    \end{subfigure}
    ~ 
    \begin{subfigure}[t]{0.4\textwidth}
        \includegraphics[width=\textwidth]{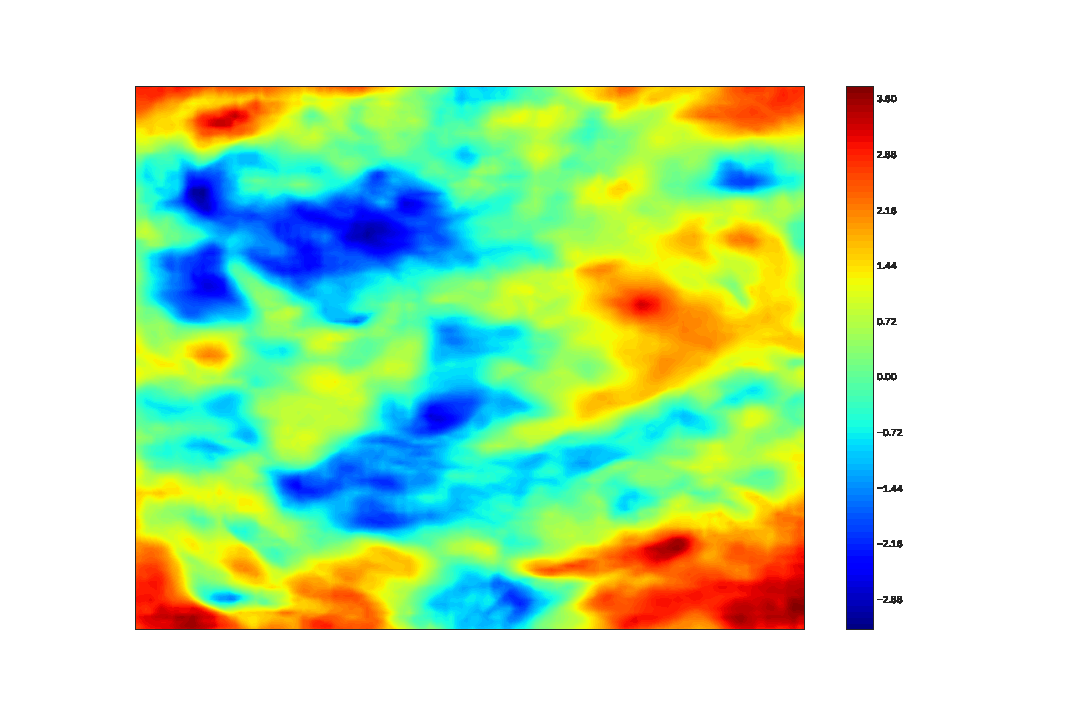}
        \caption{$V$ velocity}
        \label{modelgraphics:V}
    \end{subfigure}
    ~ 
    \\
    \begin{subfigure}[t]{0.4\textwidth}
        \includegraphics[width=\textwidth]{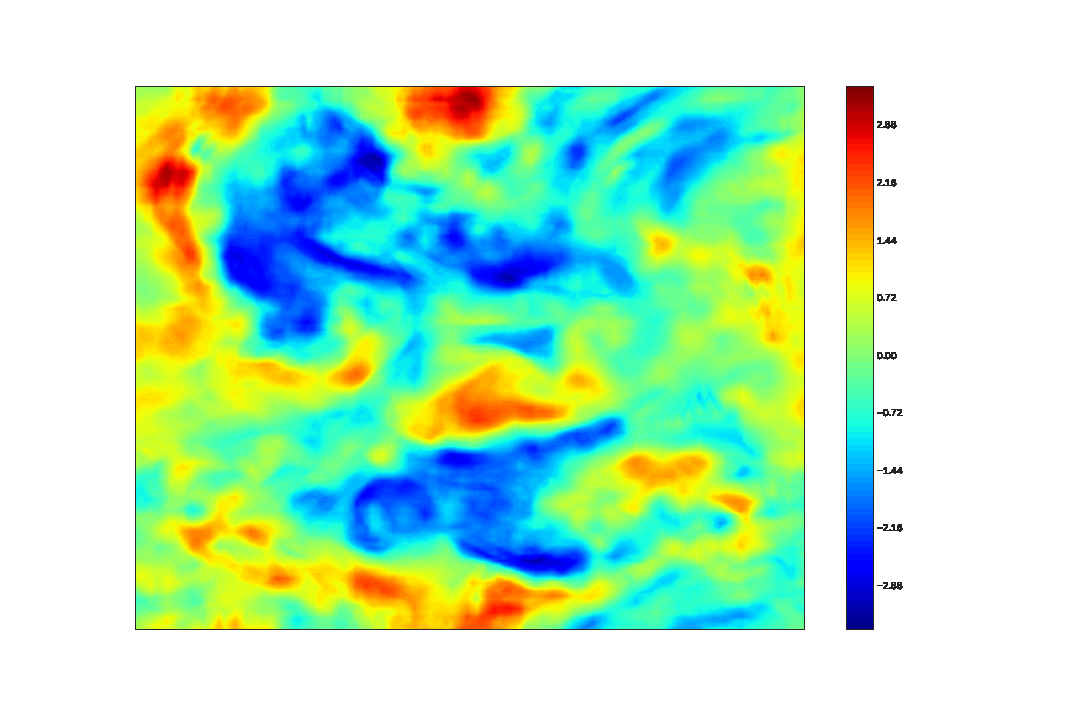}
        \caption{$W$ velocity}
        \label{modelgraphics:W}
    \end{subfigure}
    \caption{$3\mathrm{D}$ Flow realizations of HIT predicted by CC-LSTM Neural Network: $2\mathrm{D}$ slices}\label{modelgraphics}
\end{figure}
\begin{figure}
    \centering
    \begin{subfigure}[t]{0.4\textwidth}
        \includegraphics[width=\textwidth]{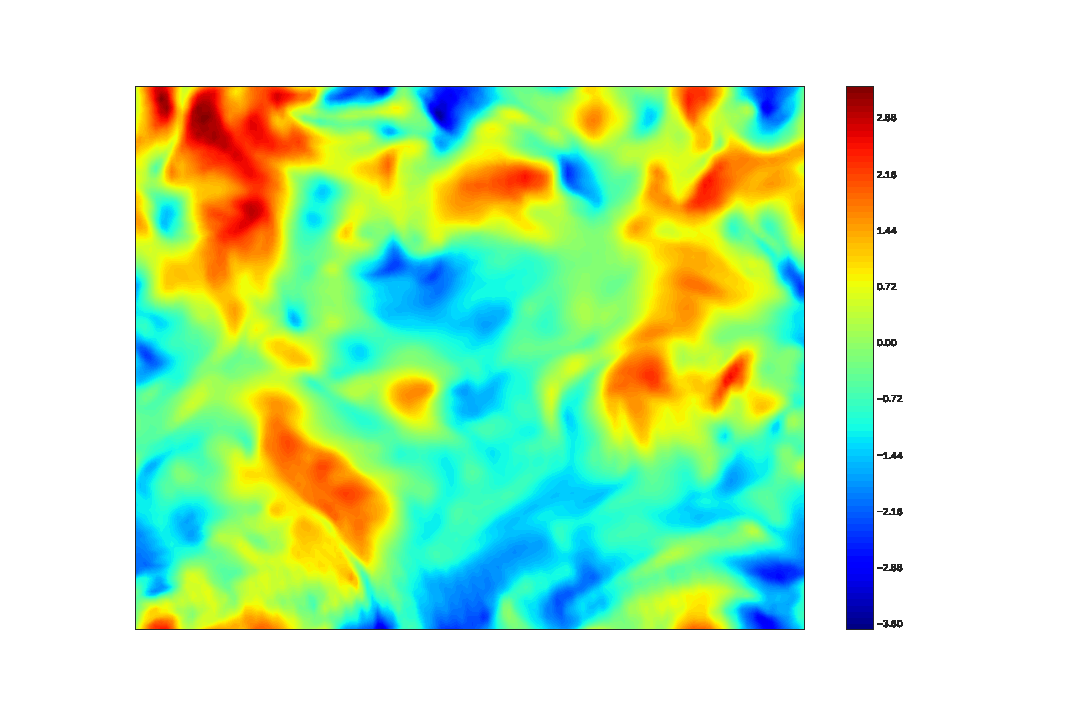}
        \caption{$U$ velocity}
        \label{dns:U}
    \end{subfigure}
    ~ 
    \begin{subfigure}[t]{0.4\textwidth}
        \includegraphics[width=\textwidth]{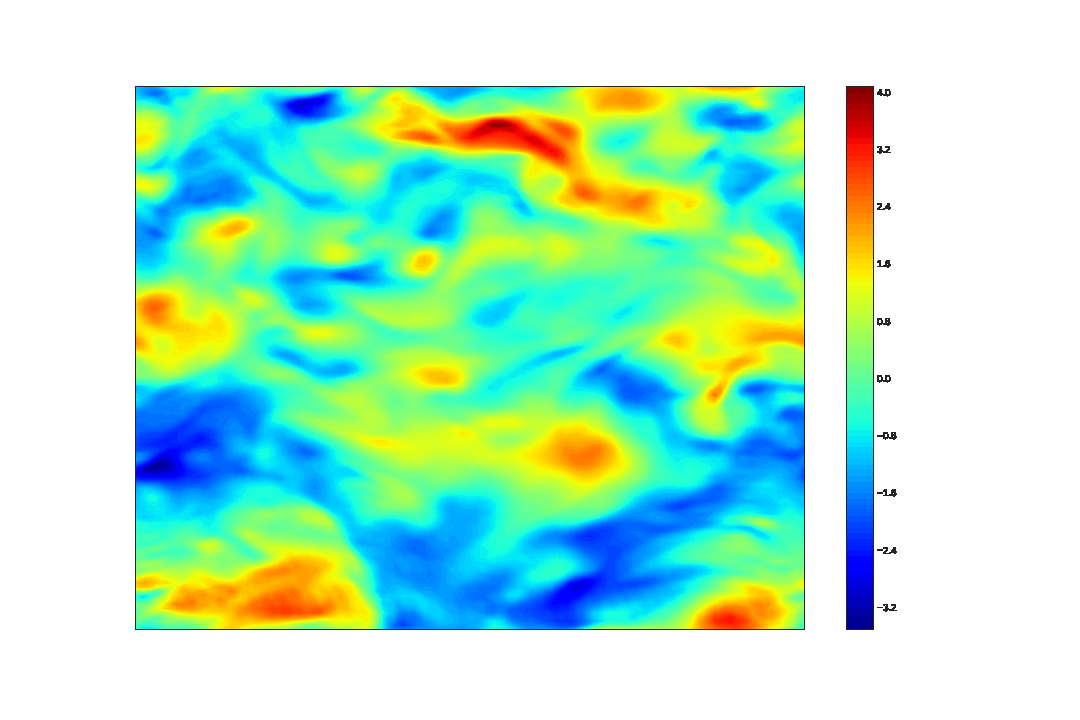}
        \caption{$V$ velocity}
        \label{dns:V}
    \end{subfigure}
    ~ 
    \\
    \begin{subfigure}[t]{0.4\textwidth}
        \includegraphics[width=\textwidth]{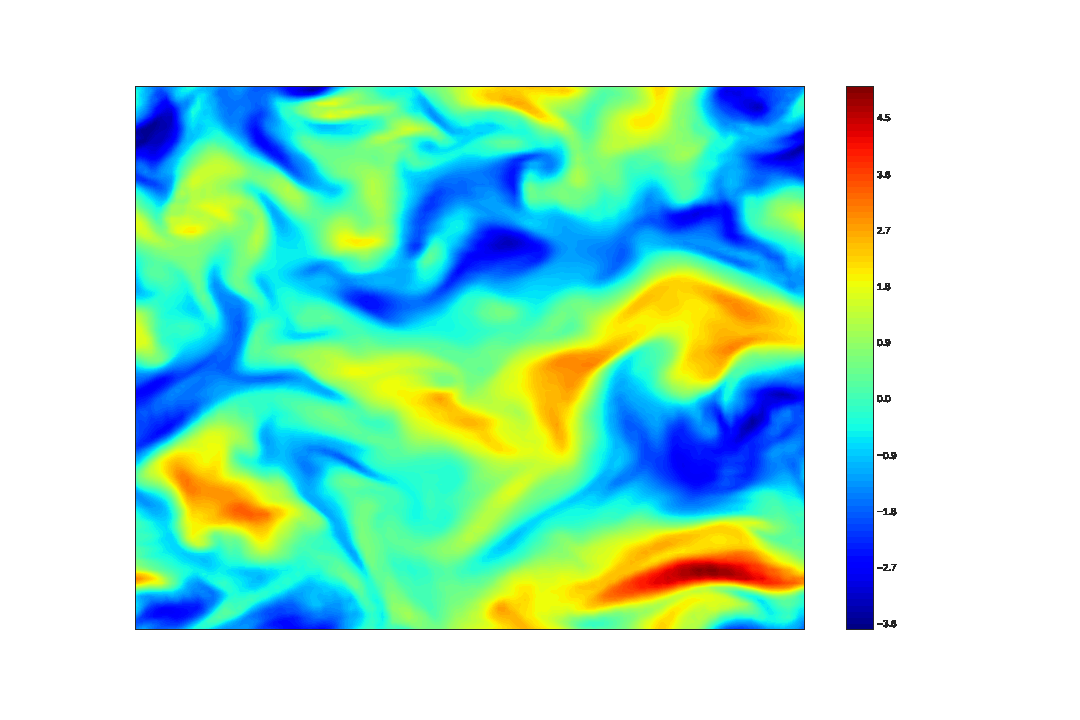}
        \caption{$W$ velocity}
        \label{dns:W}
    \end{subfigure}
    \caption{$3\mathrm{D}$ Flow realizations of HIT from DNS dataset: $2\mathrm{D}$ slices}\label{dnsgraphics}
\end{figure}
The HIT dataset described in Section~\ref{datadescription} is used as training data, with $20$ snapshots containing $U$,$V$ and $W$ components of velocity. The CAE dimensionality reduction generates latent spaces with a compression ratio of $z=125$. The learned model is used to generate predictions as described in the schematic in Fig.~\ref{cclstm-schematic}. The model generates $k$ snapshots, which are then fed back into the model as input to generate the next $k$ snapshots. Thus, this `cyclic' prediction can generate infinitely long sequences, which are ideal to study the statistics of stationary predictions and their long term stability. The realizations generated by the trained CC-LSTM model are shown in Fig.~\ref{modelgraphics}, while those from the DNS are shown in Fig.~\ref{dnsgraphics}, at a eddy turnover time of $tau=4$. While the predictions and DNS are in $3\mathrm{D}$, a randomly chosen $2\mathrm{D}$ slice is shown in the figures. It is immediately apparent that the CC-LSTM network has learned the attractor dynamics extremely well, leading to flow realizations with realistic multiscale features virtually indistinguishable from DNS. While it is often tempting in the deep learning literature to claim the success of a model from convincing visual comparisons, it is important not to ignore the scientific nature of the datasets, as those used in this work. As such, any understanding of the efficacy of a deep learning model can only come from intensive physics based accuracy metrics which are considerably different from the accuracy metrics employed in learning models. Standard accuracy metrics used in machine learning (such as mean squared error) are strictly numerical and do not provide a clear insight into the various physical features of the flow that may or may not have been captured. Therefore, in this work the accuracy of the learned attractor is quantified via statistical tests of turbulence physics. The statistical tests outlined in Section~\ref{diagnostics} have been designed to analyze the model predictions of turbulence features through Kolmogorov spectra, small scale statistics and flow morphology. These test give different perspectives into how the model performs, and directly link it to turbulence physics of interest. 
\begin{figure}
    \centering
    \begin{subfigure}[t]{0.2\textwidth}
        \includegraphics[width=\textwidth]{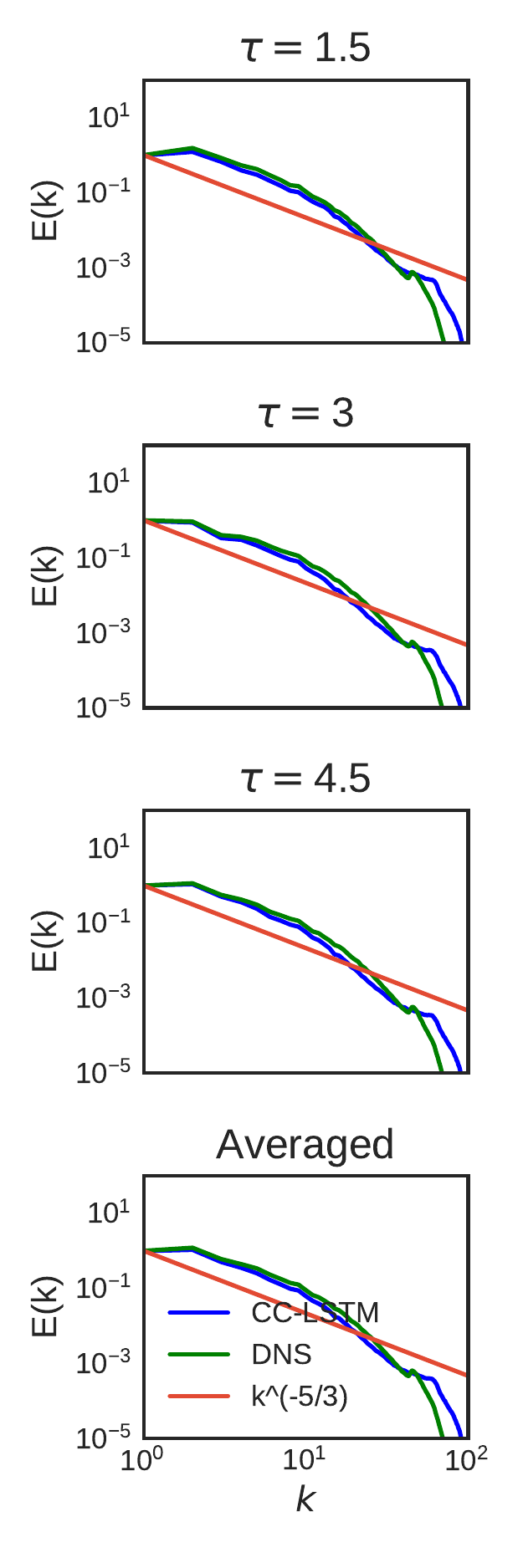}
        \caption{Energy Spectra}
        \label{HIT_CCLSTM:spec}
    \end{subfigure}
    ~ 
    \begin{subfigure}[t]{0.2\textwidth}
        \includegraphics[width=\textwidth]{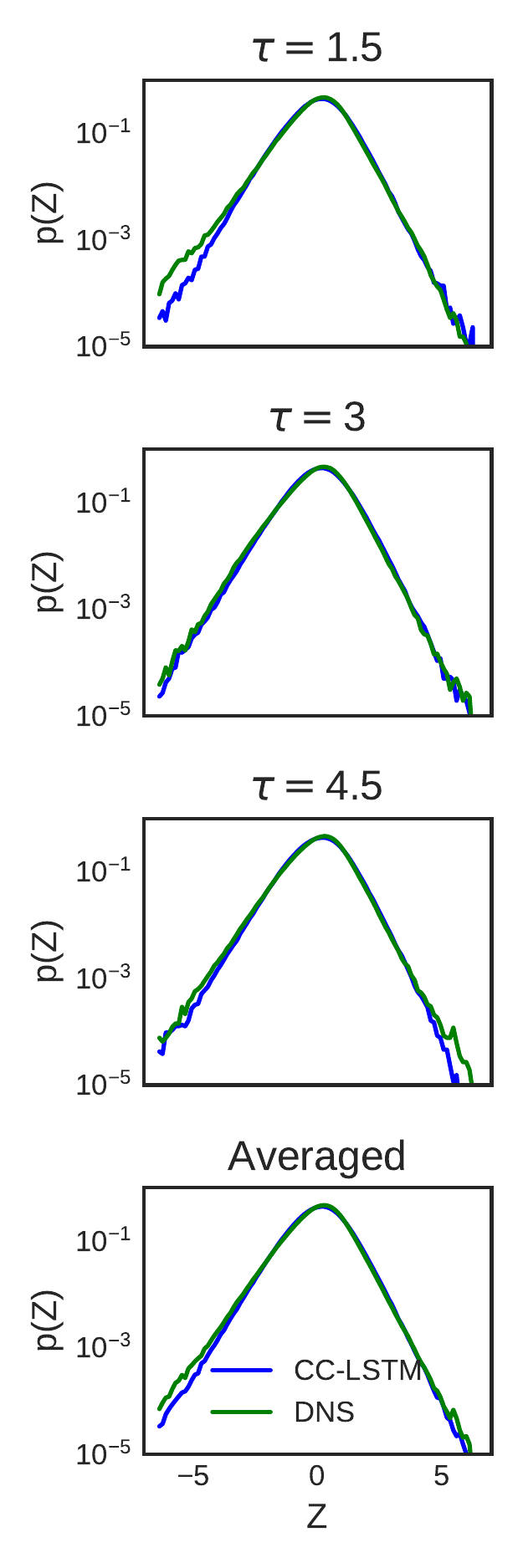}
        \caption{PDF of Velocity Gradients}
        \label{HIT_CCLSTM:PDF}
    \end{subfigure}
    ~ 
    \\
    \begin{subfigure}[t]{0.4\textwidth}
        \includegraphics[width=\textwidth]{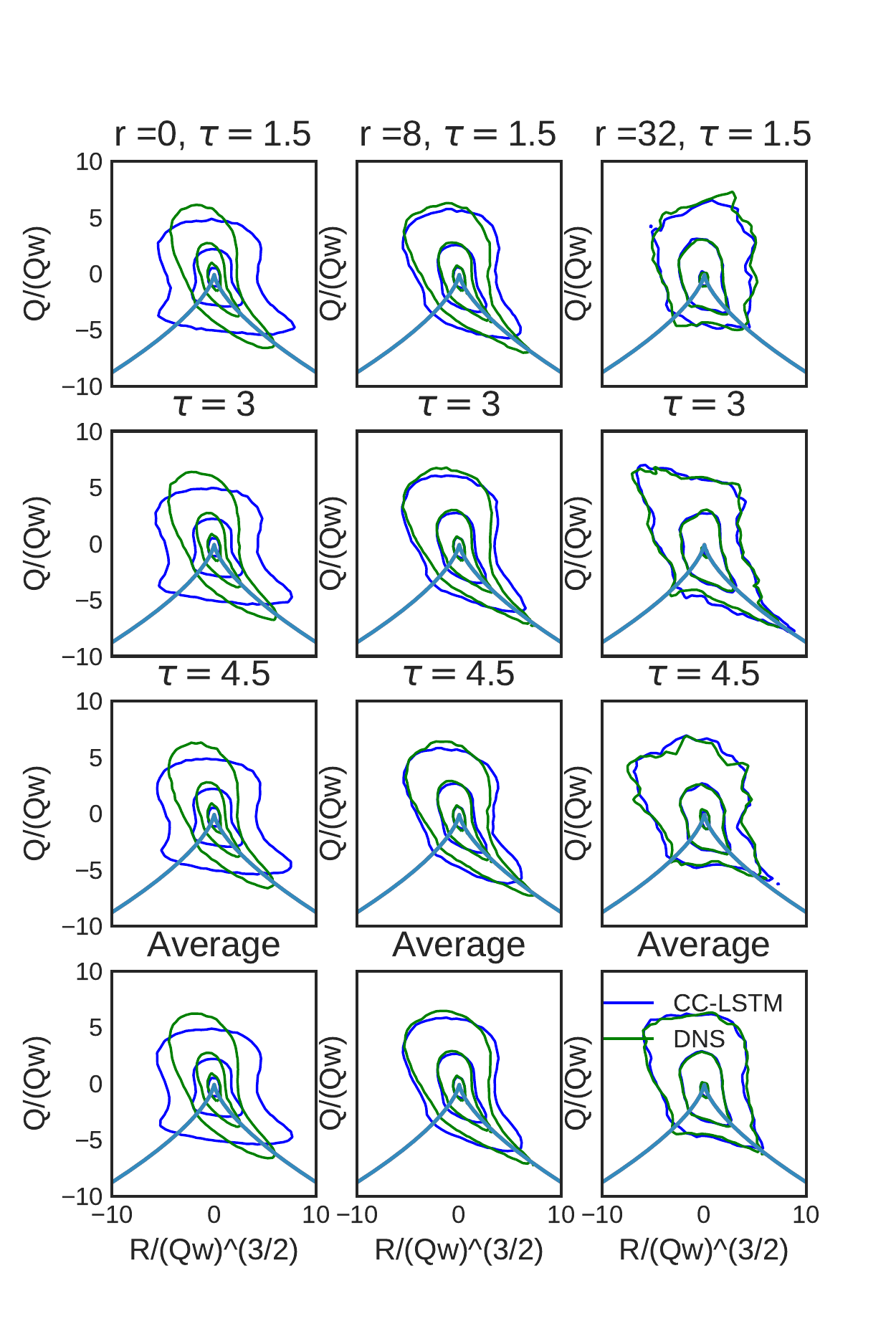}
        \caption{$Q-R$ plane at various coarse-grained scales}
        \label{HIT_CCLSTM:qr}
    \end{subfigure}
    \caption{Statistics of Predicted Snapshots of HIT from CC-LSTM ROM}\label{HIT_CCLSTM}
\end{figure}
The results of the diagnostics tests for model predictions are shown in Fig.~\ref{HIT_CCLSTM}. Since the cyclic prediction produces long term predictions, we use the snapshots every $1.5$ eddy turnover times $\tau$ apart to compute statistics. This would not only provide the accuracy of turbulence physics predicted at every snapshot, but also if the accuracy varies with increasing the prediction horizon. We present results of statistical tests on 3 different instants, $\tau=1.5$ apart. Finally, an averaged diagnostic is computed from these 3 snapshots and shown at the bottom. Specifically, we have $3$ diagnostics tests a) The Kolmogorov energy spectra, b) PDFs of velocity gradient, which exemplifies the small scale intermittent features, and c) The $Q-R$ plane structure. While the first two tests are integrated quantities, the $Q-R$ plane is a test of flow morphology. The $r=0,8,32$ show the comparison between flow morphologies of DNS and model predictions at various scales. Small scales are represented in $r=0$, inertial/mid-range scales in $r=8$ and large scales in $r=32$. Now we are in a position to critically analyze the model accuracy.

It is immediately apparent from all 3 tests that the large scales are modeled extremely well by CC-LSTM. This can be seen in the low wavenumber energy spectra, and the $r=32$ plots in the $Q-R$ plane. Furthermore, it is exceptional that the large dynamics are also retained temporally, as seen at $\tau = 1.5,3,4.5$. This is the strength of using a learning approach based on LSTM networks, which explicitly learn temporal dynamics, as will be seen in the subsequent sections. The inertial scale dynamics in $r=8$ are also captured in both the energy spectrum and velocity gradient PDFs, albeit with minor discrepancies. The $Q-R$ plane shows that the inertial-scale range by and large captures the structure of DNS scales, with some differences. As with the large scales, the temporal dynamics in $r=8$ are also well resolved. Finally, we direct our attention to the small scale behavior shown in high wavenumber Kolmogorov spectra and $r=0$ in the $Q-R$ plane. The previous section outlined the intricacies of dimensionality reduction with CAE, where the choice was made to avoid capturing all scales in favor of minimizing computing costs of a CC-LSTM ROM. Furthermore, we made connections between convolutional kernel strides and turbulent scales, and hypothesized that our choice of $\beta = 2$ would lead to small scales being washed out. The small scale behavior captured in all $3$ diagnostics consistently indicate that the small scales have been indeed neglected, as intended from our CAE design. The energy spectra shows discrepancies in the high wavenumber region, the intermittency PDFs also show deviations at the tails corresponding to smaller scales. The final test is the small scale flow morphology, which fails to capture any of the DNS small scale dynamics. Therefore, there appears to be some merit in our connections between the design parameters of Convolutional neural networks and the multiscale nature of turbulence. We note that one of the key criticisms of machine learning based physics models is the lack of physical interpretation in the design of deep neural networks. From our results, we hope to show that physics-inspired interpretation and design of deep learning models is certainly possible, and a fruitful area of future investigation for the community.

Finally, it is important to evaluate the long-term stability of the cyclic-predictions from the CC-LSTM model. To this end, we compute the diagnostic tests at various instants from the model prediction to see any potential degradation in quality. Much like numerical schemes, unbounded errors can accumulate with every iteration and eventually become unstable, leading to erroneous predictions. We find that the CC-LSTM model is able to predict snapshots upto $\tau =7$ without loss of stability and degradation in flow statistics. This is particularly impressive, since the training was done on only $\tau = 1.25$ snapshots; further indicating that CC-LSTM model can suitably approximate the underlying high-dimensional attractor from only a few snapshots.

\subsection{Homogeneous Isotropic Turbulence with Passive Scalars}
The ScalarHIT dataset is similar to the previous case, since the velocity components are HIT. However there are two key differences which help understand the behavior of CC-LSTM even further: a) The addition of two passive scalars with predetermined statistics~\cite{daniel2018reaction}, and b) A compression ratio of $z=20$ is used for the latent space, which is much lower than HIT. The  RA  method  mathematically  models  fast  chemical  reactions  occurring between hypothetical reactants identified in a mixed fluid state. Therefore this case was chosen as a test bed to explore the potential of CC-LSTM to model reacting flow components, in addition to only advection (exemplified by the HIT case). Another motivation is to study if a lower compression ratio results in a better modeling accuracy at the inertial and larger velocity scales, while selectively eliminating only the smaller scales. 
\begin{figure}
    \centering
    \begin{subfigure}[t]{0.2\textwidth}
        \includegraphics[width=\textwidth]{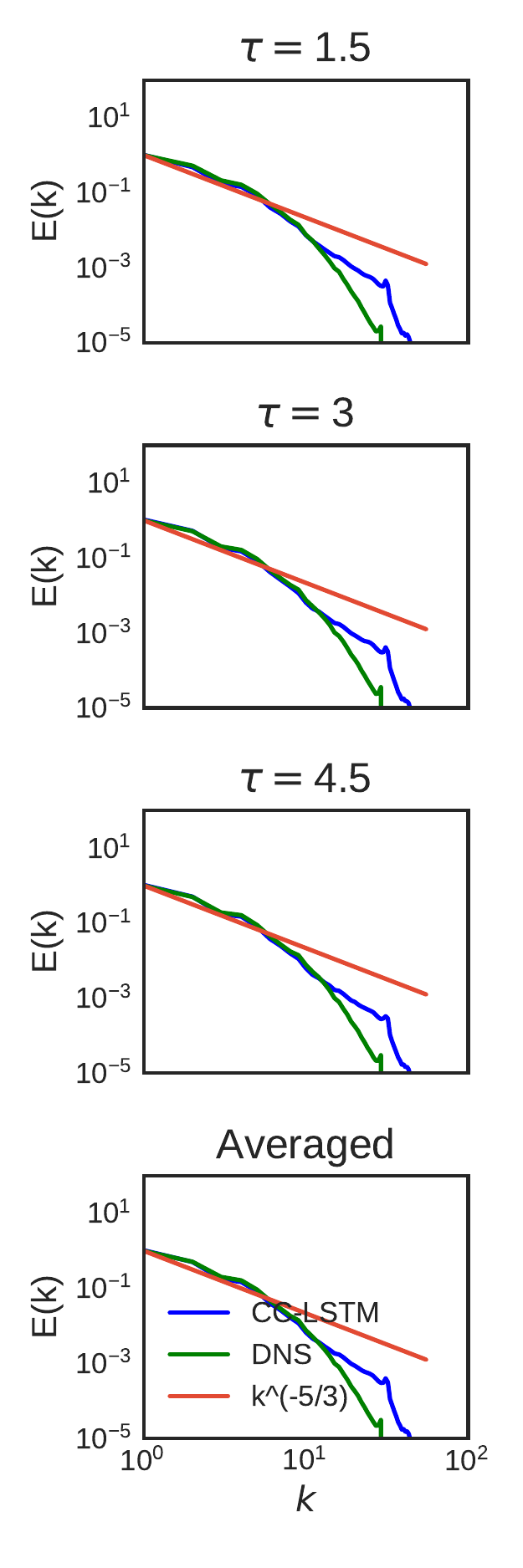}
        \caption{Energy Spectra}
        \label{ScalarHIT_CCLSTM_flow:spec}
    \end{subfigure}
    ~ 
    \begin{subfigure}[t]{0.2\textwidth}
        \includegraphics[width=\textwidth]{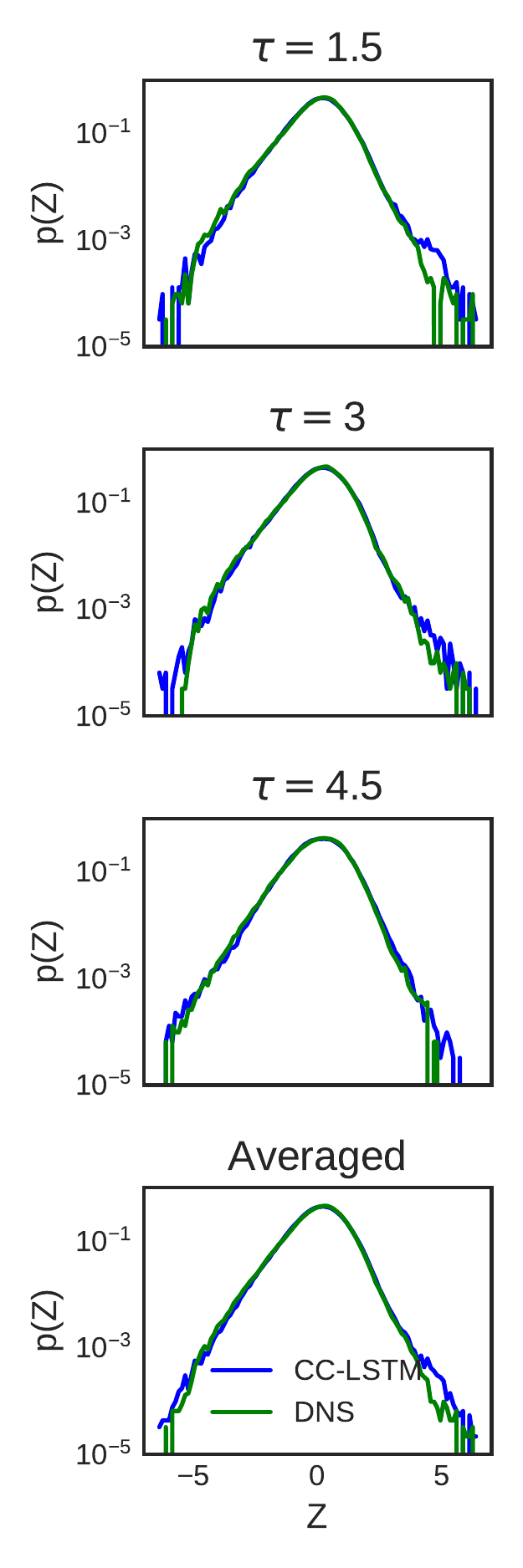}
        \caption{PDF of Velocity Gradients}
        \label{ScalarHIT_CCLSTM_flow:PDF}
    \end{subfigure}
    \begin{subfigure}[t]{0.4\textwidth}
        \includegraphics[width=\textwidth]{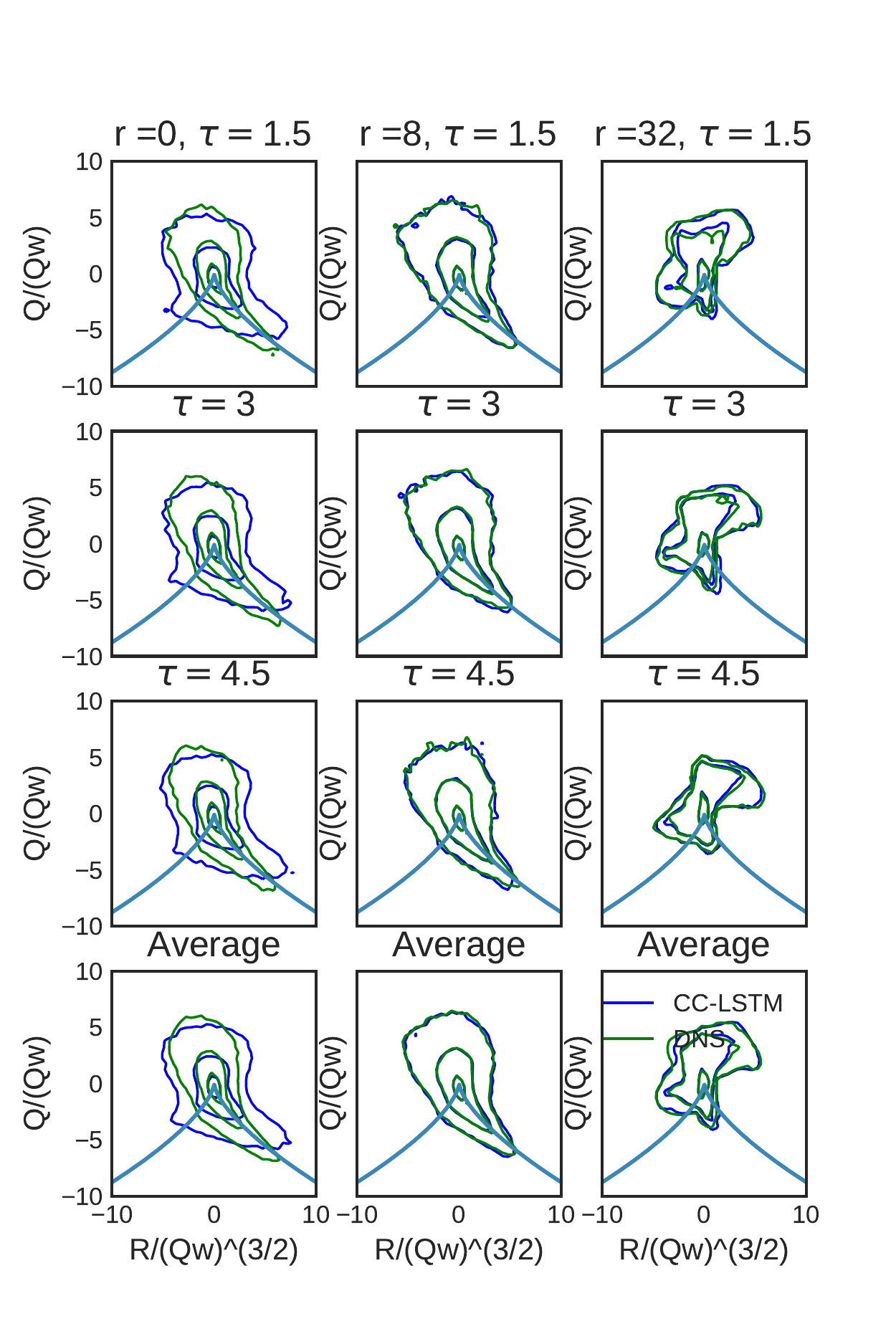}
        \caption{$Q-R$ plane at various coarse-grained scales}
        \label{ScalarHIT_CCLSTM_flow:qr}
    \end{subfigure}
    \caption{Statistics of Predicted Snapshots of ScalarHIT from CC-LSTM ROM}\label{ScalarHIT_CCLSTM_flow}
\end{figure}
As before, we again use $20$ snapshots to train, but with $5$ variables - three velocity components and two passive scalars $\phi_{1}$ and $\phi_{2}$ per snapshot. For the velocity components, we use the same diagnostic tests as HIT and the results are shown in Fig.~\ref{ScalarHIT_CCLSTM_flow}. The figure shows clear improvements compared to HIT diagnostics in Fig.~\ref{HIT_CCLSTM}. The energy spectra in ScalarHIT shows an excellent match for both the high and mid-wavenumbers, corresponding to large and inertial scales. In particular, the error appears to be isolated primarily in the small scales. This behavior is also observed from the intermittency PDFs and where the error is concentrated exclusively in the tails. Finally, the flow morphology from the $Q-R$ planes clearly show superior statistics in the large and inertial range scales, with small scale error similar to that seen in HIT. A key difference is the lack of discrepancies in $r=8$ that was seen in HIT. We now compare the statistics of the passive scalars. The RA forcing methodology in Ref.~\cite{daniel2018reaction} specifies fixed bounds for the evolution of each of the passive scalars, and their PDFs. The goal here is to explore how CC-LSTM models these scalars, and if the small scale effects observed before also manifest here.
\begin{figure}
    \centering
    \begin{subfigure}[t]{0.4\textwidth}
        \includegraphics[width=\textwidth]{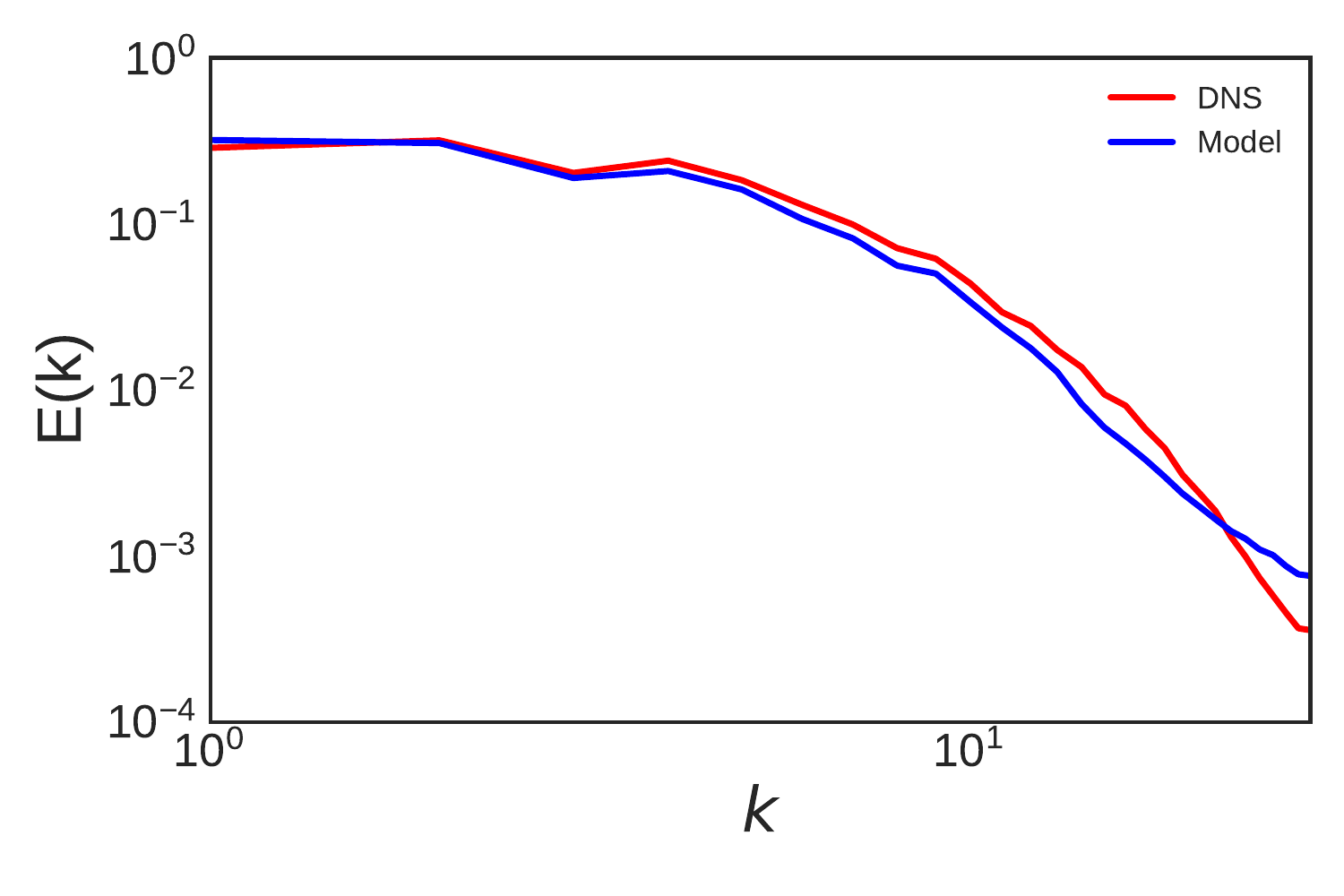}
        \caption{Energy Spectra for $\phi_1$}
        \label{ScalarHIT_CCLSTM_scalars:specY1}
    \end{subfigure}
    ~ 
    \begin{subfigure}[t]{0.4\textwidth}
        \includegraphics[width=\textwidth]{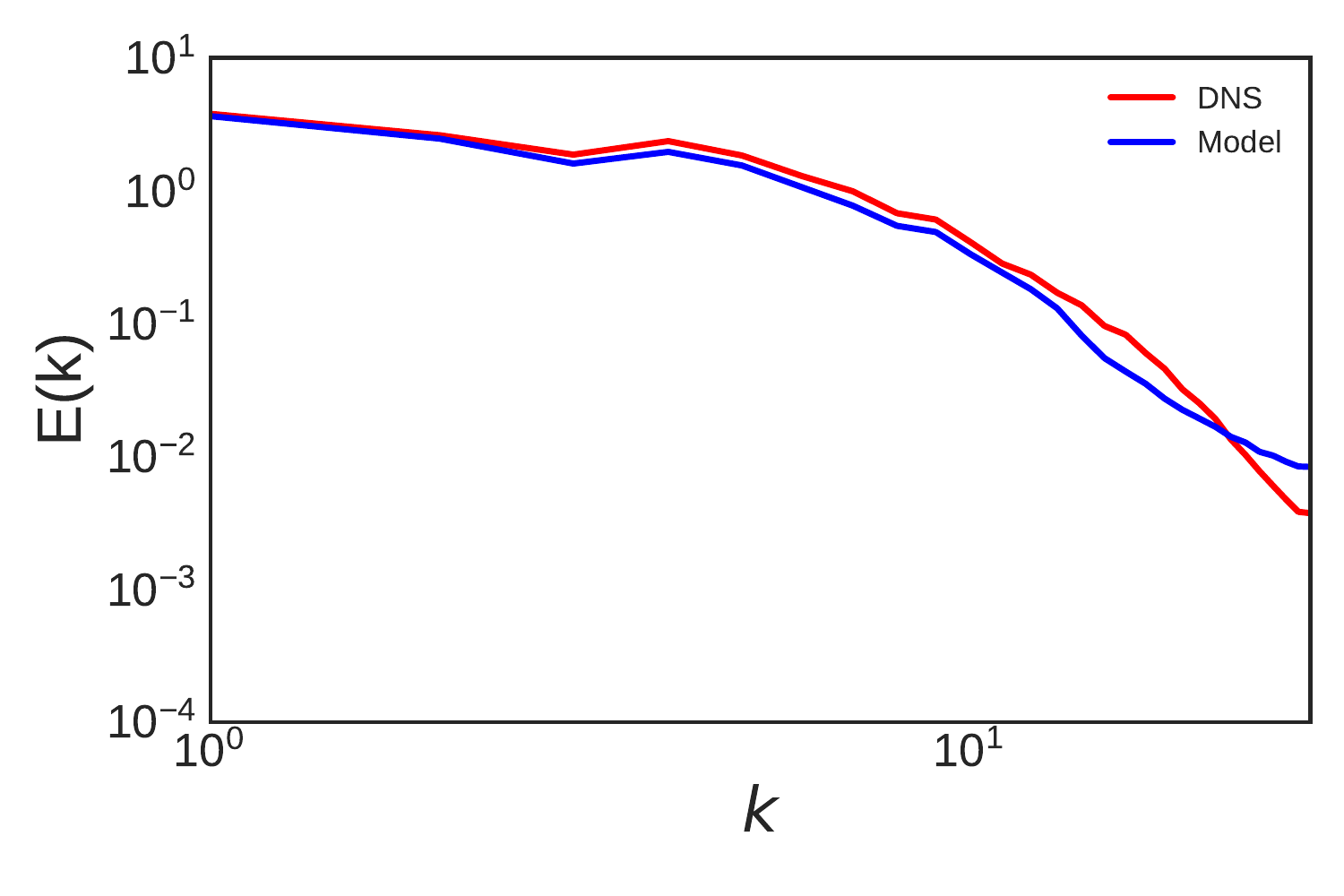}
        \caption{Energy Spectra for $\phi_2$}
        \label{ScalarHIT_CCLSTM_scalars:specY2}
    \end{subfigure}
    \\
    \begin{subfigure}[t]{0.4\textwidth}
        \includegraphics[width=\textwidth]{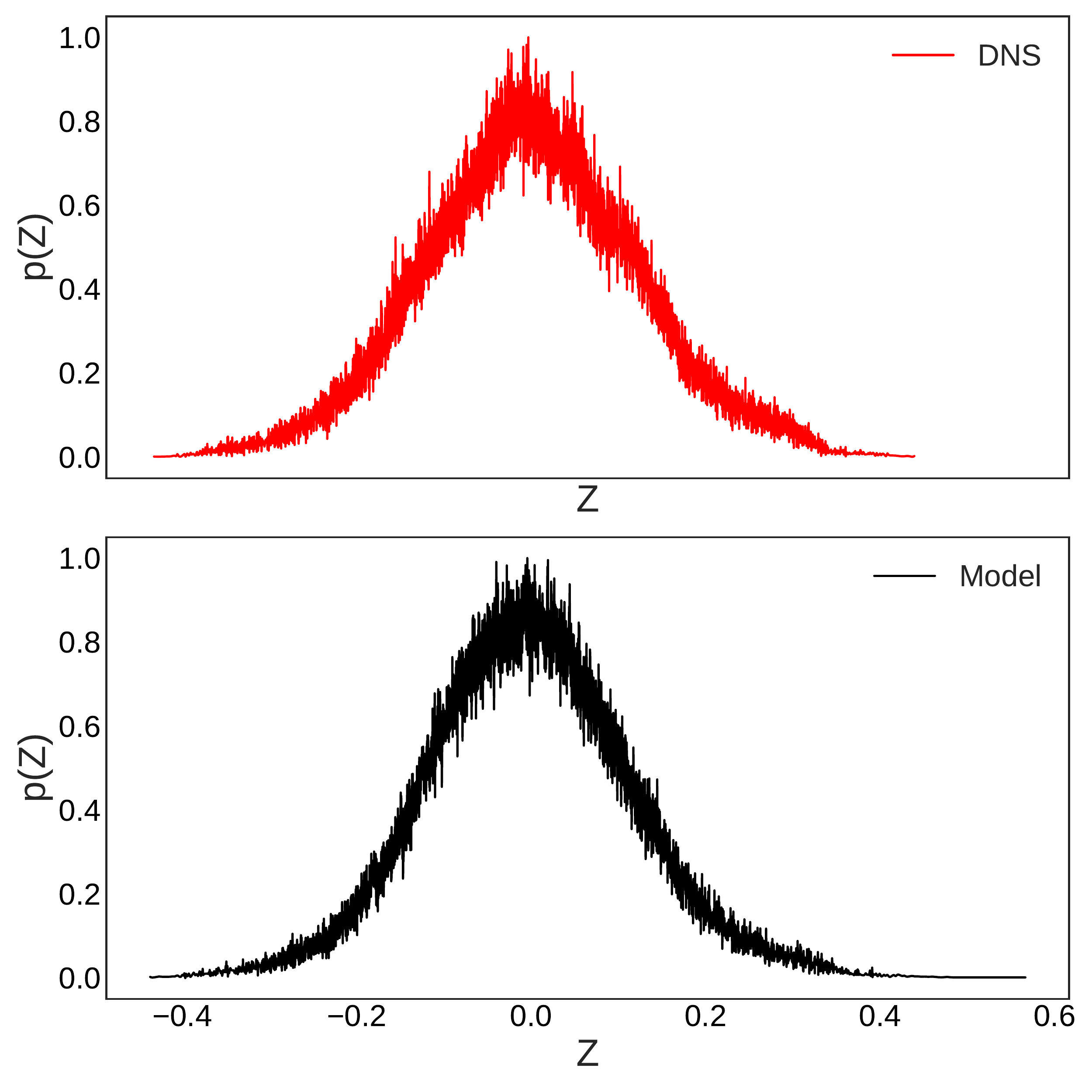}
        \caption{PDF for $\phi_1$}
        \label{ScalarHIT_CCLSTM_scalars:PDFY1}
    \end{subfigure}
    \begin{subfigure}[t]{0.4\textwidth}
        \includegraphics[width=\textwidth]{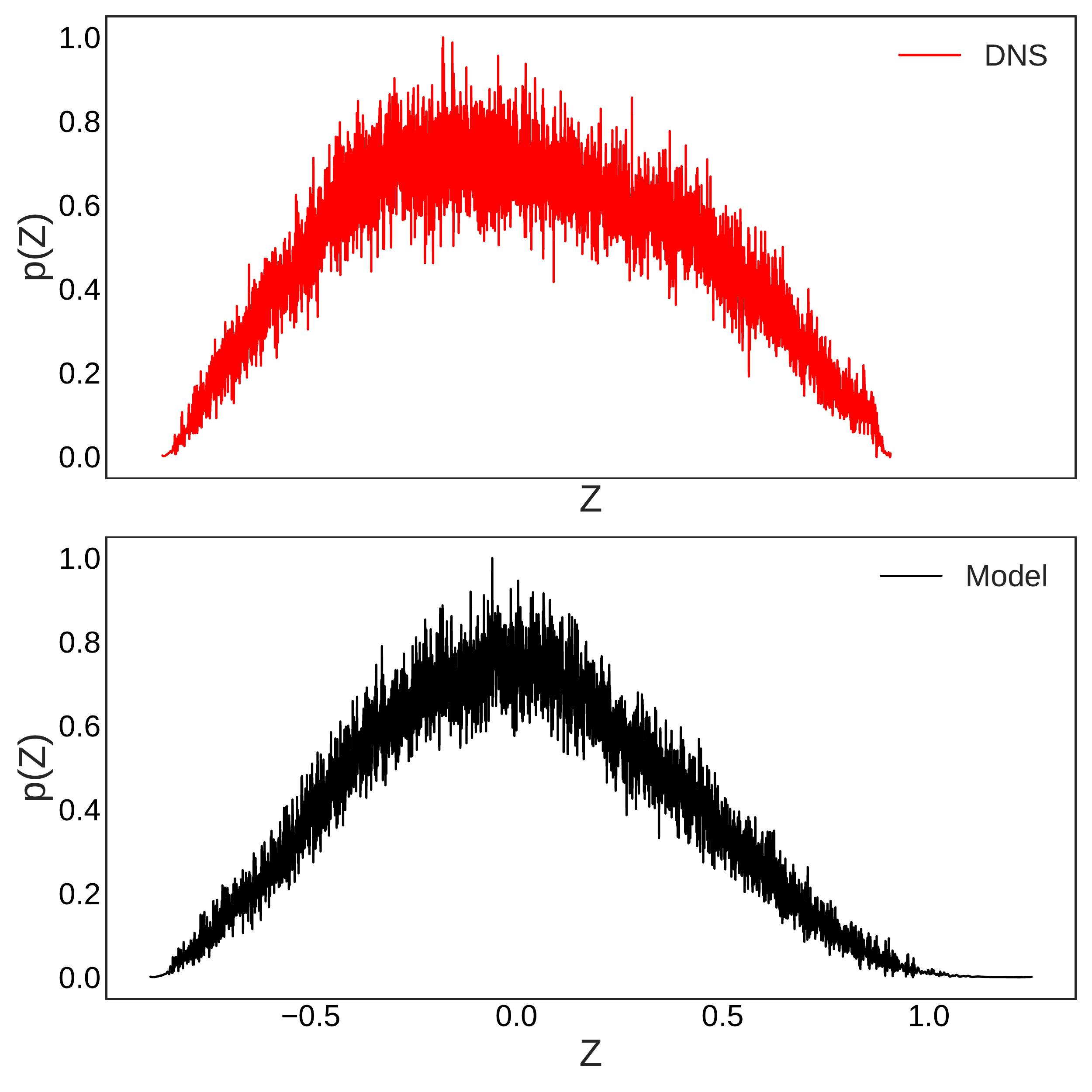}
        \caption{PDF for $\phi_2$}
        \label{ScalarHIT_CCLSTM_scalars:PDFY2}
    \end{subfigure}
    \caption{Statistics of passive scalars in ScalarHIT from CC-LSTM ROM}\label{ScalarHIT_CCLSTM_scalars}
\end{figure}
Two diagnostic tests are used for $\phi_{1}$ and $\phi_{2}$ - PDF of the statistics and Energy spectra. The PDFs for each of the two scalars, and their corresponding CC-LSTM predictions are shown in Fig.~\ref{ScalarHIT_CCLSTM_scalars}. It is seen that the model predicts the PDF for $\phi_{1}$ extremely well, whereas it is less accurate for $\phi_{2}$, with visible discrepancies around $+/-0.5$. 
However, a more interesting behavior is observed at the tails for both PDFs. In the RA forcing method, both scalars are enforced with strict upper and lower bounds, with $\phi_{1}$ bounded in the interval $[-0.5,0.5]$ and $\phi_{2}$ bounded $[-1,1]$. The PDF from the model predictions for $\phi_{1}$ show a minute, but unambiguous unbounded behavior past $0.5$. Likewise, a similar unbounded PDF is also visible in $\phi_{2}$ past $1.0$. We now focus on the spectra of the two passive scalars; which are shown in Fig.~\ref{ScalarHIT_CCLSTM_scalars:specY1} and Fig.~\ref{ScalarHIT_CCLSTM_scalars:specY2}. The clearer picture emerges, as it is seen that the high wavenumber features are completely neglected by the compression scheme, and other wavenumbers are predicted reasonably well. Overall, the CC-LSTM scheme shows capability in modeling additional quantities of interest apart from the advection terms, and physics-inspired CAE design appears to influence the predicted scales as desired.

\section{Influence of LSTM Cells}
\label{sec:lstminfluence}
\subsection{Effect of LSTM cells in CC-LSTM}
We explored in detail the behavior of the Convolutional networks as autoencoders in Section~\ref{CAEdetail}. In this section we delve deeper into architecture of the CC-LSTM. The Convolutional LSTM cell fuses the spatial correlations captured by the Convolution operator and has connections to physics of multiscale dynamics, with the temporal dynamics encoded in the LSTM Cells. Since the data is statistically stationary, and convolutional networks are able to effectively compress and reproduce its statistics, how useful are LSTMs in the architecture? To address this, we start with a falsifiable hypothesis: For stationary dynamics, we presume LSTMs have minimal influence on predictive accuracy of CC-LSTM models.

\begin{figure}
    \centering
    \begin{subfigure}[t]{0.2\textwidth}
        \includegraphics[width=\textwidth]{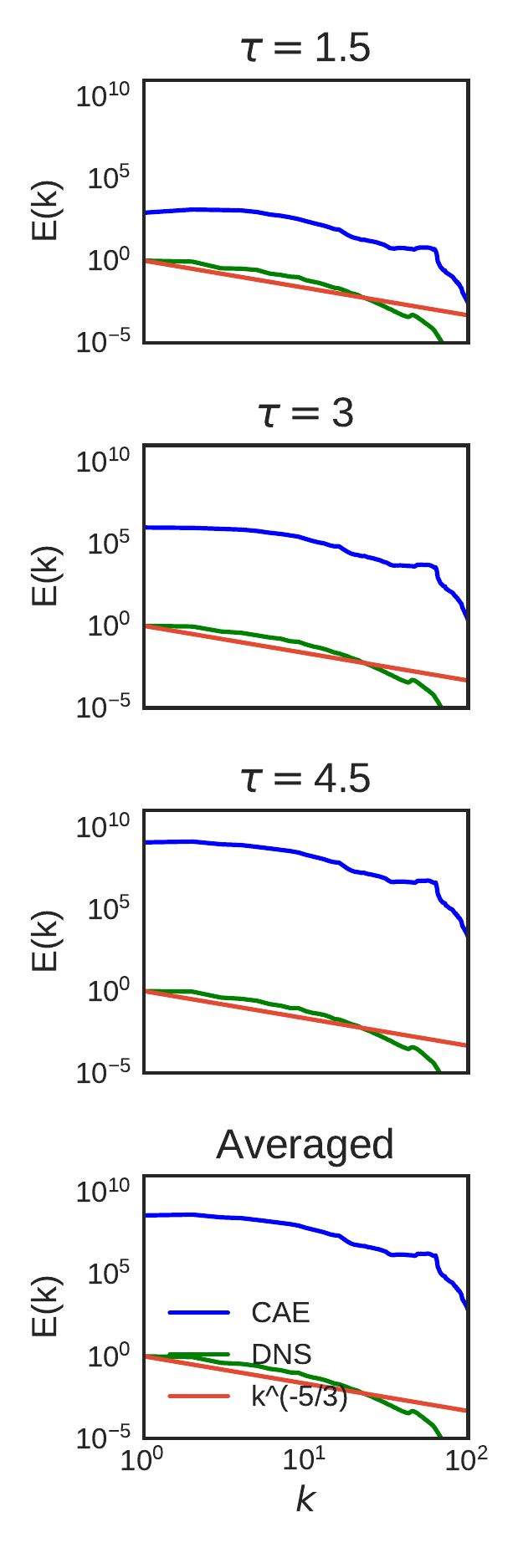}
        \caption{Energy Spectra}
        \label{CAEfailure:spec}
    \end{subfigure}
    ~ 
    \begin{subfigure}[t]{0.2\textwidth}
        \includegraphics[width=\textwidth]{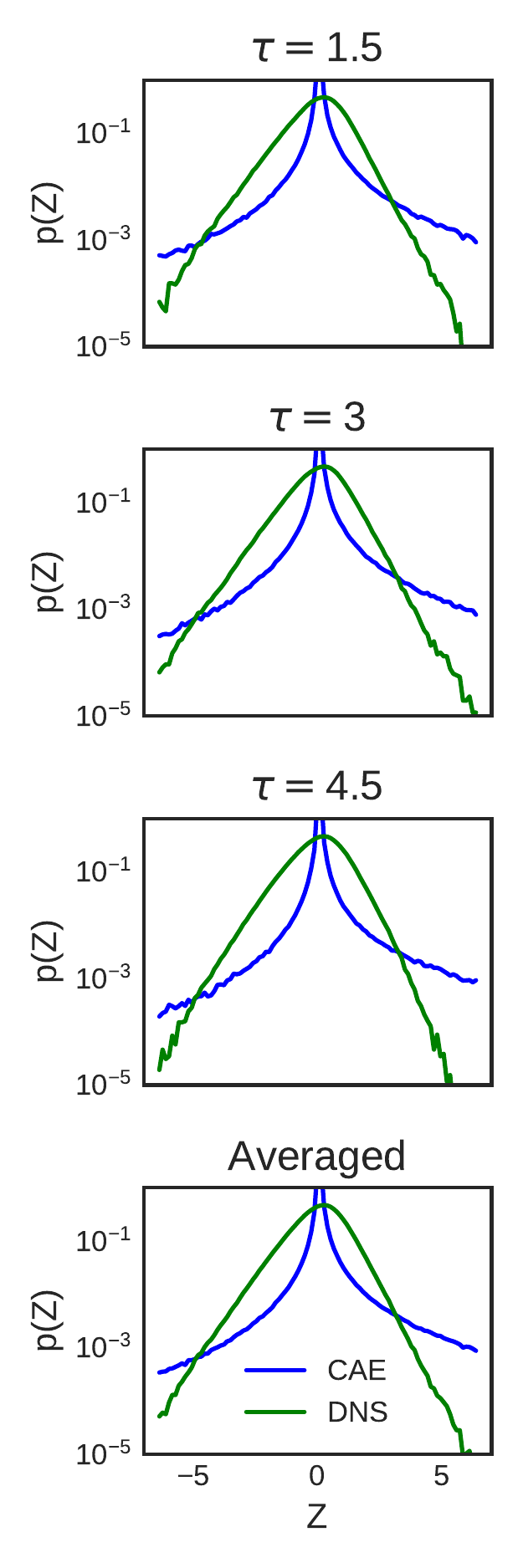}
        \caption{PDF of Velocity Gradients}
        \label{CAEfailure:PDF}
    \end{subfigure}
    ~ 
    \\
    \begin{subfigure}[t]{0.4\textwidth}
        \includegraphics[width=\textwidth]{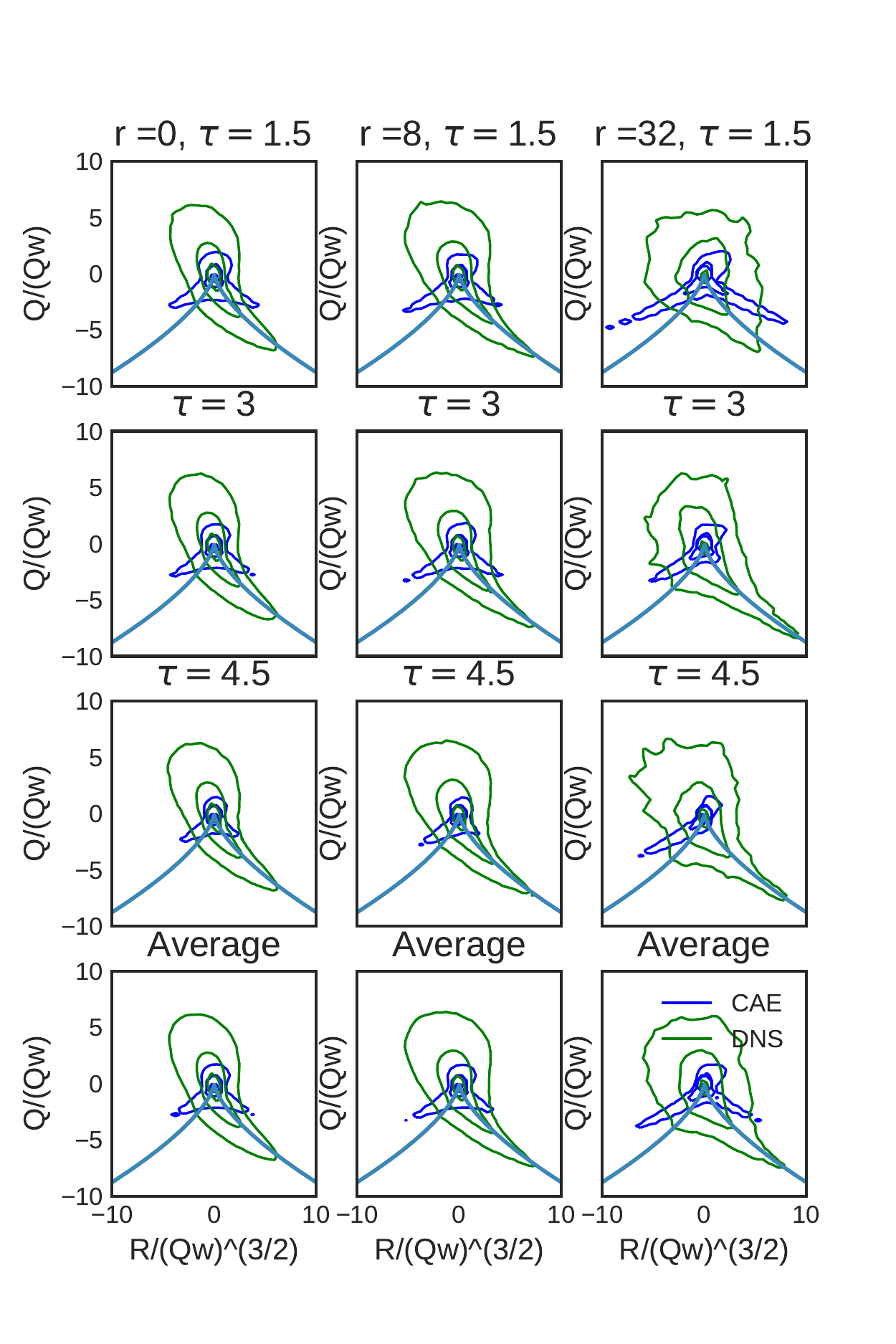}
        \caption{$Q-R$ plane at various coarse-grained scales}
        \label{CAEfailure:qr}
    \end{subfigure}
    \caption{Statistics of CAE predictions (without LSTM)}\label{CAEfailure}
\end{figure}

To test this hypothesis, we train the HIT dataset again with the same parameters, but with the ConvLSTM cells in the CC-LSTM architecture (Fig.~\ref{cclstm-schematic}) removed. The network now uses the CAE to compress the snapshots, decompress and feed them as a cyclic input back to the CAE,  to wash out the network initialization and operate in fully predictive mode. The CAE can now be used to make long term predictions of the HIT attractor dynamics. We perform the same diagnostic tests as Section~\ref{results} on a) The predicted sequences early in time, and b) predicted sequnces later in time. The goal here is to explore if the predictions are accurate and/or stable over long term, like the CC-LSTM results demonstrated in Figs~\ref{HIT_CCLSTM} and \ref{ScalarHIT_CCLSTM_flow}. We compute the statistics for the CAE predictions at the same eddy turnover times as CC-LSTM, and the results are shown in Fig.~\ref{CAEfailure}. The statistics of the energy spectra are consistently off by several orders of magnitude, with the intermittent velocity PDFs also showing erroneous profiles. To further understand the severity of these discrepancies, the $Q-R$ planes are studied in Fig.~\ref{CAEfailure:qr}. The $Q-R$ planes show a complete breakdown of predicted flow structures at all scales, in contrast to the consistent predictions generated by CC-LSTM in Section~\ref{results}. Therefore, the results lead us to conclusively make the claim that LSTM cells indeed have a significant role in learning the attractor dynamics, even in cases when the dataset is stationary.

\subsection{Effect of LSTM temporal history length}
\label{k_variation}

\begin{figure}
    \centering
    \begin{subfigure}[t]{0.2\textwidth}
        \includegraphics[width=\textwidth]{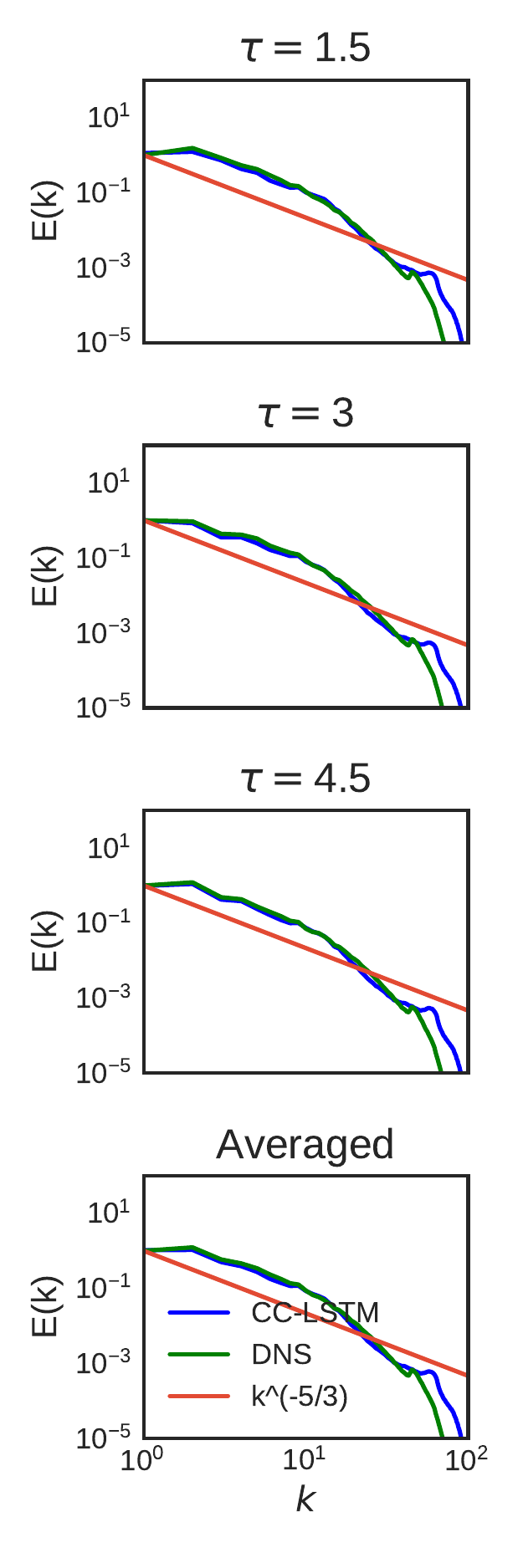}
        \caption{Energy Spectra}
        \label{kvar1_CCLSTM:spec}
    \end{subfigure}
    ~ 
    \begin{subfigure}[t]{0.2\textwidth}
        \includegraphics[width=\textwidth]{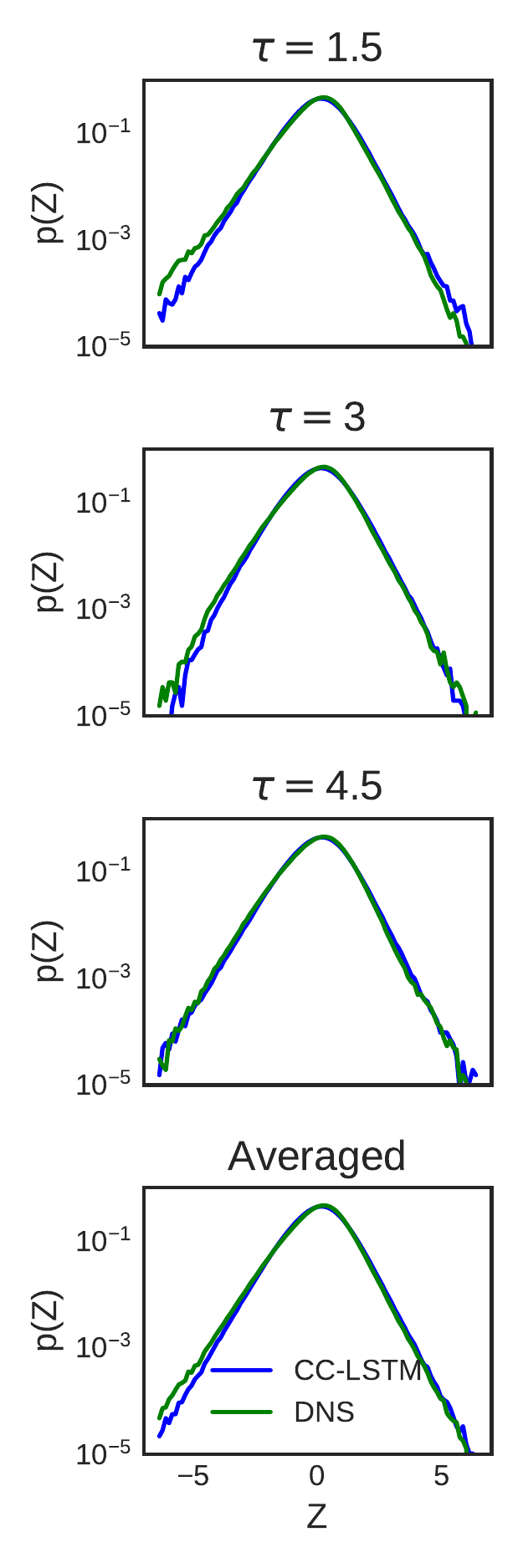}
        \caption{PDF of Velocity Gradients}
        \label{kvar1_CCLSTM:PDF}
    \end{subfigure}
    ~ 
    \\
    \begin{subfigure}[t]{0.4\textwidth}
        \includegraphics[width=\textwidth]{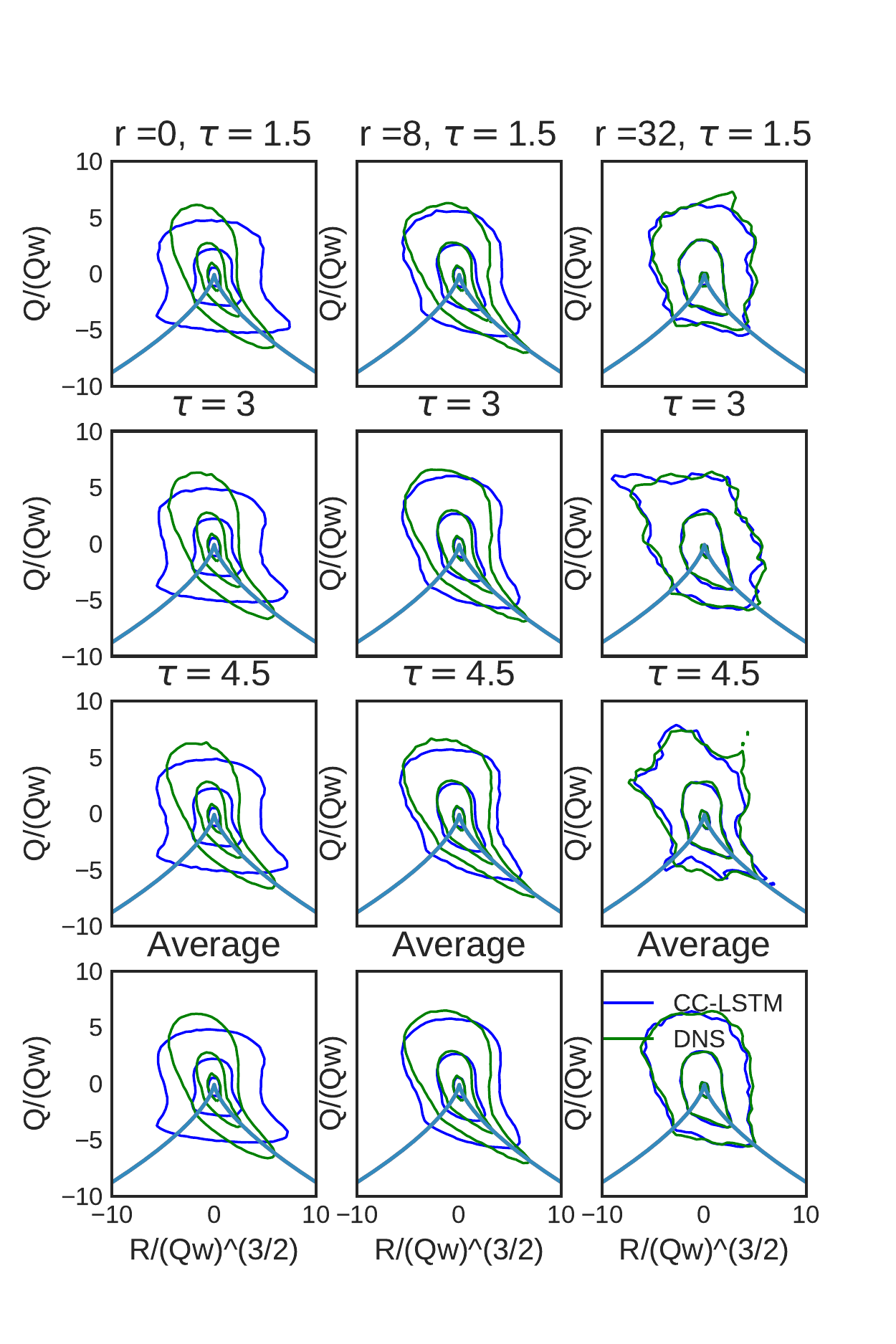}
        \caption{$Q-R$ plane at various coarse-grained scales}
        \label{kvar1_CCLSTM:qr}
    \end{subfigure}
    \caption{Statistics of Predicted Snapshots of HIT from CC-LSTM ROM: $k=1$}\label{kvar1_CCLSTM}
\end{figure}

\begin{figure}
    \centering
    \begin{subfigure}[t]{0.2\textwidth}
        \includegraphics[width=\textwidth]{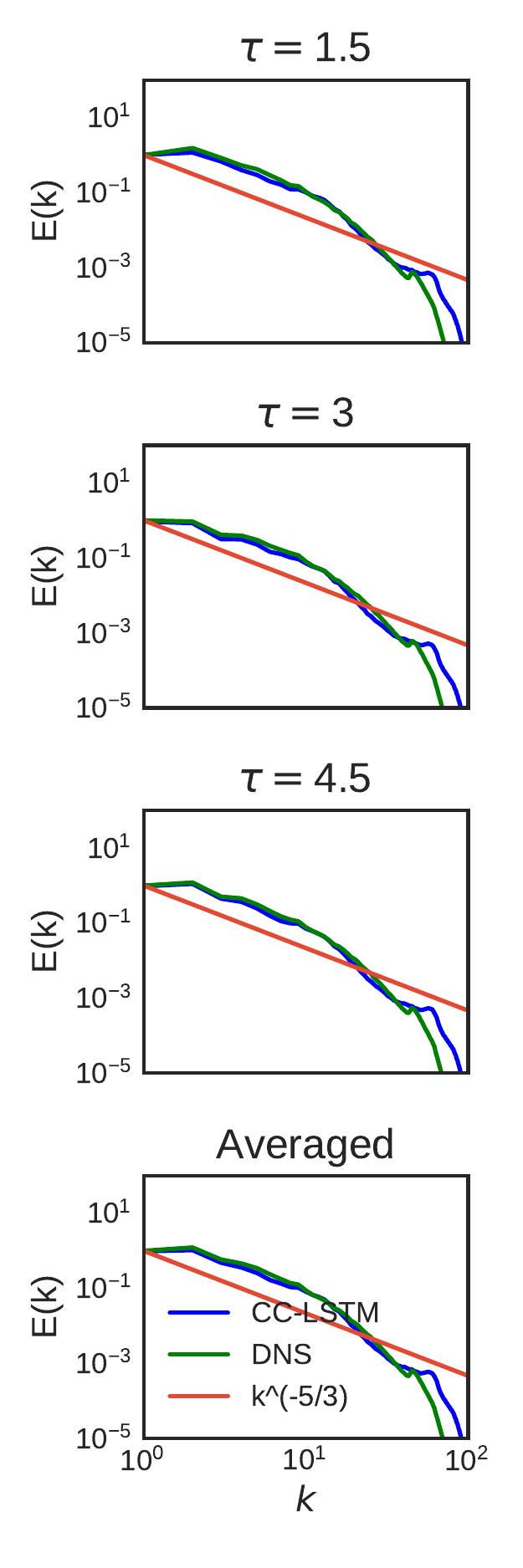}
        \caption{Energy Spectra}
        \label{kvar2_CCLSTM:spec}
    \end{subfigure}
    ~ 
    \begin{subfigure}[t]{0.2\textwidth}
        \includegraphics[width=\textwidth]{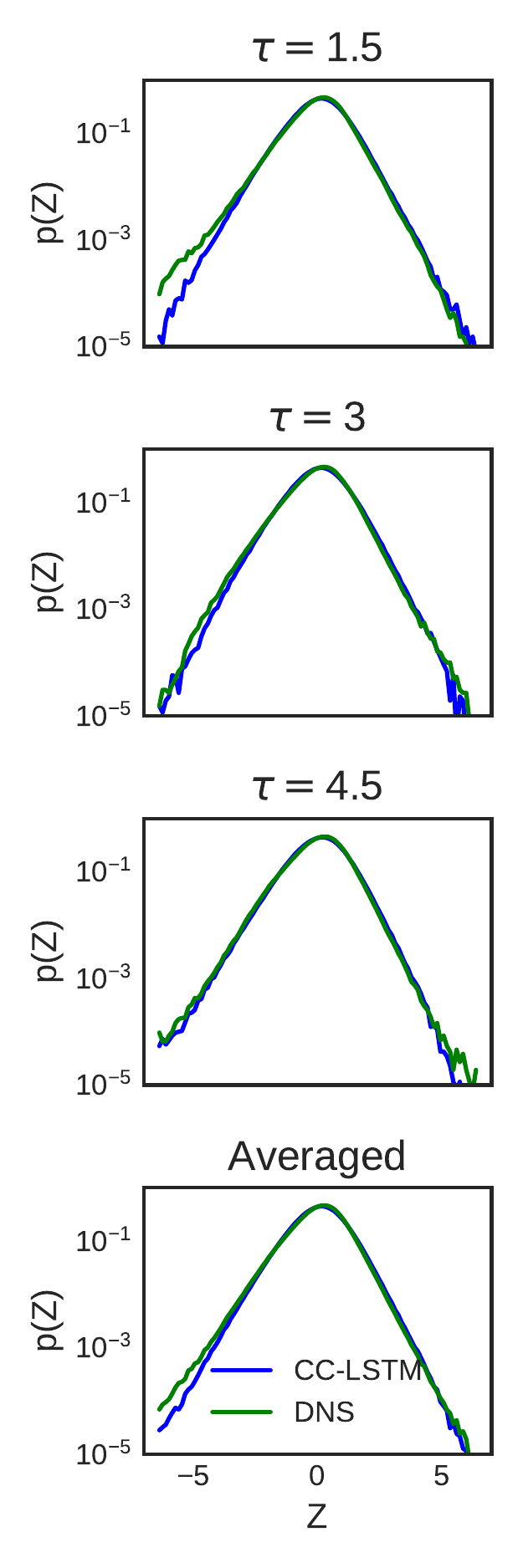}
        \caption{PDF of Velocity Gradients}
        \label{kvar2_CCLSTM:PDF}
    \end{subfigure}
    ~ 
    \\
    \begin{subfigure}[t]{0.4\textwidth}
        \includegraphics[width=\textwidth]{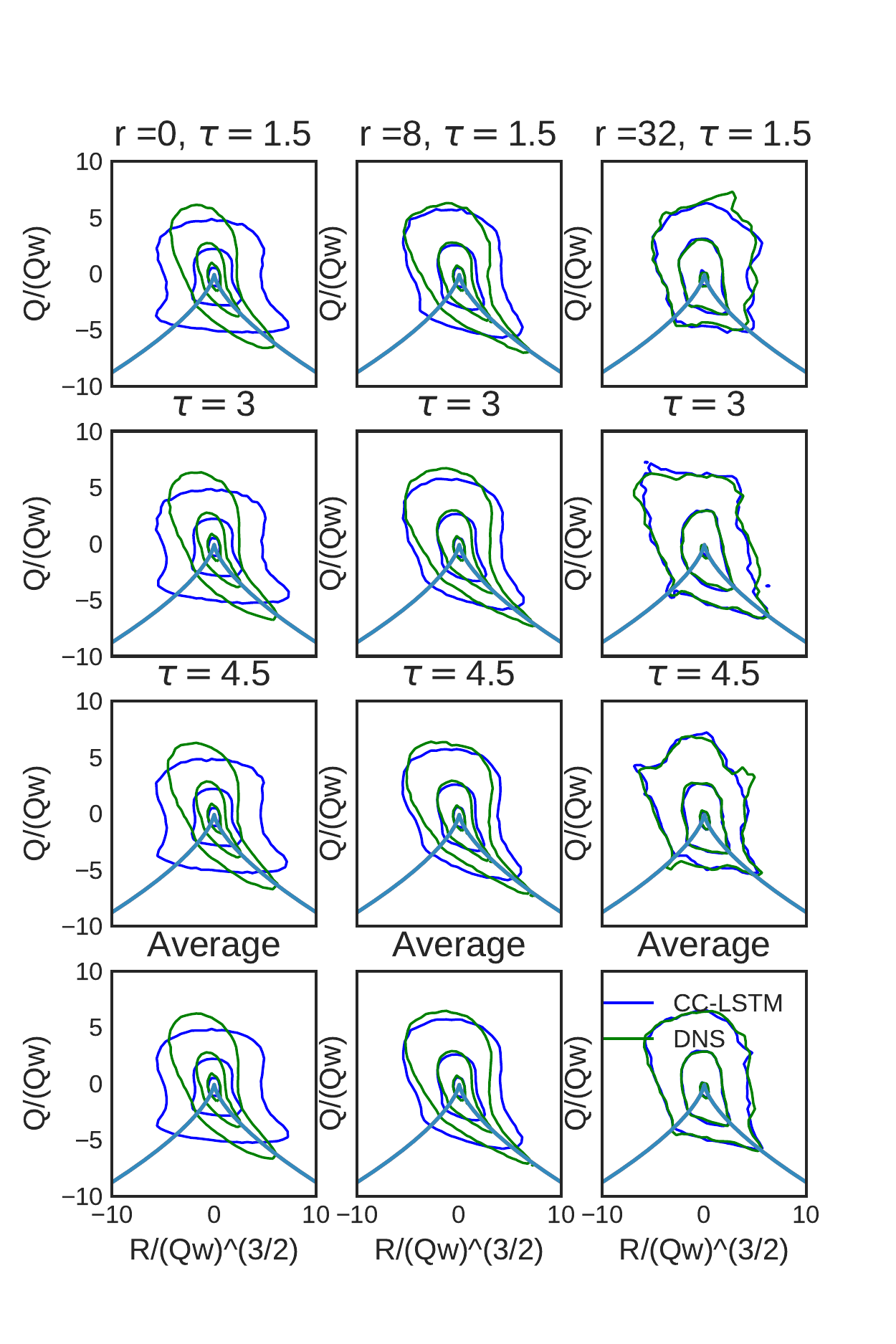}
        \caption{$Q-R$ plane at various coarse-grained scales}
        \label{kvar2_CCLSTM:qr}
    \end{subfigure}
    \caption{Statistics of Predicted Snapshots of HIT from CC-LSTM ROM: $k=2$}\label{kvar2_CCLSTM}
\end{figure}

Recurrent networks are specialized for sequential modeling, and hence the sequential dependence of any prediction on previous values in the sequence is a crucial factor in both machine learning and physics.
This is encoded by an important hyper-parameter in LSTM (and other recurrent networks) represented by $k$ in~\ref{cclstm-schematic}. The value of $k$ determines the temporal map that the LSTM networks learn, responsible for the "memory" in LSTM networks. In mathematical form,

\begin{equation}
    y_{i} = f(y_{i-1,..,i-1-k})
\end{equation}

where $y$ is a variable in a sequence, such that $y$ at any point in the sequence is a function of its $k$ previous realizations. The choice of $k$ has important ramifications on the training performance, computational cost and accuracy of an LSTM-based model. In this work, the sequence is temporal and hence $k$ corresponds to the previous time instants, and we had chosen $k=3$ for the results shown in Section~\ref{results}. In practice, $k$ is estimated based on our prior knowledge of the dynamics in the dataset, such as any known periodic frequencies and correlations. Consequently, in the absence of clear insight into these dynamics an approximate $k$ has to be chosen. In such events, care has to be taken to ensure the model predictions are reproducible for a range of $k$ and consistent with the physics of the problem, and not primarily artifacts of any specific $k$.

In both the HIT and ScalarHIT cases, the flow is driven primarily by advection and exhibits temporally stationary statistics. The stationarity makes it difficult to explicitly choose a $k$ based on the dynamics. Therefore, we repeat the model training in Section~\ref{results} with different choices of $k$, ranging from $k=1-3$ for the HIT dataset. The diagnostic tests for the three $k$ values are performed, with the $Q-R$ planes being of most interest since flow morphology is very sensitive to minor changes in predictive quality, compared to integrated quantities like energy spectra.

The $Q-R$ planes computed from long term CC-LSTM predictions for HIT with $k=1$ and $k=2$ are presented in Figs.~\ref{kvar1_CCLSTM} and  \ref{kvar2_CCLSTM} for comparison. Only the HIT dataset is considered here for brevity, since the advection in ScalarHIT is also HIT and produced identical results. The statistics from these figures, in comparison to the $k=3$ results in Fig.~\ref{HIT_CCLSTM} show the relative consistency of results for different values of $k$. While there are minor deviations in the results, the qualitative results remain remarkably consistent, especially at the various scales in the $Q-R$ plane. Overall, this exercise provides further confidence that the results obtained are not a numerical artifact of a specific $k$. Instead, these are indicative of the learned model being robust and hence less sensitive to $k$. A comment can also be made about the qualitative results holding equally well for $k=1$ and $k>1$: A $k=1$ implies purely Markovian dynamics, and hence lack of long-term memory in the system. We can hypothesize that the temporal stationarity of the flow in homogeneous, isotropic turbulence minimizes the existence of long term correlations in the flow, which can adversely impact its future realizations. Extending this hypothesis, it can be expected that CC-LSTM models for flows with non-trivial long term correlations would perform poorly for $k=1$. Though investigating this is outside the scope of the this work, it is an important avenue for future work i.e. to study the impacts of non-stationarity on CC-LSTM performance.

\section{Conclusions}
We present one of the first efforts on fully data-driven learning of the underlying attractor in three dimensional turbulence; with the goal of emulating its physics at a very low computational cost. To achieve this, we propose a framework called \textit{Compressed Convolutional LSTM} (CC-LSTM) to perform dimensionality reduction, spatio-temporal modeling and physics-based analysis of model accuracy. We show that recent developments in recurrent neural networks like Convolutional LSTM show promise in modeling multi-scale high-dimensional dynamical systems. It is also seen that dimensionality reduction is necessary to reduce the costs of training Convolutional LSTM networks for $3\mathrm{D}$ datasets. We note that Convolutional Autoencoders are a highly effective approach for dimensionality reduction by compressing the dataset into a low dimensional latent space, which is used to train the Convolutional LSTM network. Since neural networks generally suffer from lack of interpretability, we have made attempts to connect the design of convolutional autoencoder to turbulence physics. Our results imply merit in this connection and can lead to physics-inspired design of convolutional networks for turbulence. Also studied is the role of LSTM-based neural networks in the long term stability of predicted statistics, due to its strengths in capturing temporal dynamics. Due to the latent space compression, the CC-LSTM approach is also very computationally efficient, and all the networks in this work were trained on a single Nvidia 1080Ti GPU, with the overall compute resources required being $\approx 250-300$ times lower than a comparable DNS. It is also important to mention that the algorithm can be scaled up to large datasets on HPC resources, and smaller domain sizes were used in this work since the objective was a proof of concept, rather than real-world scale modeling. Such modeling efforts would be part of a future work.

In summary, this work shows the immense potential of generative deep learning approaches like CC-LSTM to turbulence and the techniques described herein can be further extended to other spatio-temporal dynamical systems. We also emphasize on the role of rigorous physics based tests to quantify the success of a deep learning model, since visual comparison as a sole metric has several pitfalls. While further work is necessary to include more physical constraints and interpretability in LSTM models, there is confidence that spatio-temporal modeling with this class of neural networks has significant applications in modeling climate phenomena, astrophysical flows and engineering systems.

\section{Appendix}\label{appendix}
In this section we review basic statistical concepts commonly used in the modern literature to analyze results of theoretical, computational and experimental studies of homogeneous isotropic incompressible turbulence in three dimensions. Combination of these concepts are used in the main part of the manuscript as a metric to analyze the accuracy of the deep learning model.

We assume that a $3\mathrm{D}$ snapshot, or its $2\mathrm{D}$ slice, or a temporal sequence of snapshots of the velocity field, ${\bm v}=(v^{(i)}({\bm r})|i=1,\cdots,3)$, is investigated. We will focus here on analysis of static correlations within the snapshots. We consider various objects of interest, e.g. correlation functions of second, third and fourth orders
\begin{eqnarray}
&&  \hspace{-1.2cm} C_2^{(i,j)}({\bf r}_1,{\bf r}_2)=\langle v^{(i)}({\bf r}_1) v^{(j)}({\bf r}_2)\rangle,\label{eq:second}\\
&&  \hspace{-1.2cm} C_3^{(i,j,k)}({\bf r}_1,{\bf r}_2,{\bf r}_3)=\langle v^{(i)}({\bf r}_1) v^{(j)}({\bf r}_2)v^{(k)}({\bf r}_3)\rangle,\label{eq:third}\\
&&  \hspace{-1.2cm} C_4^{(i,j,k,l)}({\bf r}_1,\cdots,{\bf r}_4)=\langle v^{(i)}({\bf r}_1) v^{(j)}({\bf r}_2)v^{(k)}({\bf r}_3)v^{(l)}({\bf r}_4)\rangle.\label{eq:fourth}
\end{eqnarray}
Rich tensorial structure of the correlation functions carry a lot of information. It also suggests a number of useful derived objects, each focusing on a particular feature of the turbulent flows. In particular, we may be to discuss structure functions, defined as tensorial moments of the increments between two points separated by the radius-vector ${\bf r}$:
\begin{eqnarray}
 && \hspace{-1.5cm} S_2^{(i,j)}({\bm r})=\langle (v^{(i)}({\bf r})-v^{(i)}({\bf 0}))(v^{(j)}({\bf r})-v^{(j)}({\bf 0}))\rangle, \label{eq:S2}\\
&& \hspace{-1.5cm} S_3^{(i,j,k)}({\bm r})=\langle (v^{(i)}({\bf r})-v^{(i)}({\bf 0}))(v^{(j)}({\bf r})-v^{(j)}({\bf 0}))\nonumber\\
&& \times  (v^{(k)}({\bf r})-v^{(k)}({\bf 0}))\rangle, \label{eq:S3}\\
&& \hspace{-1.5cm} S_4^{(i,j,k,l)}({\bm r})=\langle (v^{(i)}({\bf r})-v^{(i)}({\bf 0}))(v^{(j)}({\bf r})-v^{(j)}({\bf 0}))\nonumber\\
&& \times (v^{(k)}({\bf r})-v^{(k)}({\bf 0}))(v^{(l)}({\bf r})-v^{(l)}({\bf 0}))\rangle. \label{eq:S4}
\end{eqnarray}
Other objects of interest,  derived from the correlation function by spatial differentiation and then merging the points are moments of the velocity gradient tensor, $m^{(i,j)}=\nabla^{(i)} v^{(j)}$
\begin{eqnarray}
&& \hspace{-1cm} D_2^{(i_1,i_2;j_1,j_2)}=\langle m^{(i_1,j_1)}m^{(i_2,j_2)}\rangle, \label{eq:grad2}\\
&& \hspace{-1cm} D_3^{(i_1,i_2,i_3;j_1,j_2,j_3)}=\langle m^{(i_1,j_1)}m^{(i_2,j_2)}m^{(i_3,j_3)}\rangle, \label{eq:grad3}\\
&& \hspace{-1cm} D_4^{(i_1,i_2,i_3,i_4;j_1,j_2,j_3,j_4)}=\langle m^{(i_1,j_1)}m^{(i_2,j_2)}m^{(i_3,j_3)} m^{(i_3,j_3)}\rangle. \label{eq:grad4}
\end{eqnarray}
We may also be interested to study mixed objects, e.g. the so-called energy flux
\begin{eqnarray}
F({\bf r})=\langle v^{(j)}({\bf 0}) v^{(i)}({\bf r}) m^{(i,j)}({\bf r})\rangle.
\label{eq:flux}
\end{eqnarray}

The remainder of this section is organized as follows. We describe important turbulence concepts mentioned in the main part of the paper one by one,  starting from simpler ones and advancing towards more complex concepts. 

\paragraph{$4/5$ Kolmogorov law and the Energy Spectra}

Main statement of the Kolmogorov theory of turbulence (in fact, the only formally proved statement of the theory) is that asymptotically in the inertial range, i.e. at $L\gg r\gg\eta$, where $L$ is the largest, so-called energy-containing scale of turbulence and $\eta$ is the smallest scale of turbulence, so-called Kolmogorov (viscous) scale, $F(r)$ does not depend on $r$.  Moreover, the so-called $4/5$-law states for the third-order moment of the longitudinal velocity increment
\begin{eqnarray}
&& L\gg r\gg\eta:\quad S_3^{(i,j,k)}\frac{r^i r^j r^k}{r^3}=-\frac{4}{5}\varepsilon r,
\label{eq:4/5}
\end{eqnarray}
where $\varepsilon=\nu D_2^{(i,j;i,j)}/2$ is the kinetic energy dissipation also equal to the energy flux.

Self-similarity hypothesis extended from the third moment to the second moment results in the expectation that within the inertial range, $L\gg r\eta$, the second moment of velocity increment scales as, $S_2(r)\sim v_L (r/L)^{2/3}$. This feature is typically tested by plotting the energy spectra of turbulence (expressed via $S_2(r)$) in the wave vector domain, as shown in the main text. 

\paragraph{Intermittency of Velocity Gradient}
One expects moments of the velocity increment to show the following scaling behavior inside the inertial range of turbulence
\begin{eqnarray}
&& L\gg r\gg\eta:\quad S_n(r)\sim (v_L)^n\left(\frac{r}{L}\right)^{n/3-\Delta_n}
\label{eq:S_n_scaling}
\end{eqnarray}
where $L$ is the energy containing (largest) scale of turbulence, $\eta\sim (\nu/v_L)^{3/4}L^{1/4}$ is the Kolmogorov (viscous) scale, $v_L$ is the typical velocity fluctuations at the energy containing scale, $\nu$ is the viscosity coefficient and $\Delta_n$ is the so-called anomalous scaling. $\Delta_3=0$, and $\Delta_n$ is a growing function of $n$. Consistently with Eq.~(\ref{eq:S_n_scaling}), estimation of the moments of the velocity gradient results in
\begin{eqnarray}
&& D_n\sim \frac{S_n(\eta)}{\eta^n},
\label{eq:D_n_scaling}
\end{eqnarray}
where dependence of the $r/L$- and $v_L$- independent pre-factors in both Eq.~(\ref{eq:S_n_scaling}) and Eq.~(\ref{eq:D_n_scaling}) on $n$ is ignored. Intermittency (extreme non-Gaussianity) of turbulence is expressed the strongest in Eq.~(\ref{eq:D_n_scaling}).

\paragraph{Statistics of coarse-grained velocity gradients: $Q-R$ plane.}
The second and third invariant of the velocity gradient tensor are often important quantities for analysis, since they exhibit alignment with the intermediate eigenvector/eigenvalue of the strain rate tensor. The joint probability density function of these velocity gradient tensor invariants are denoted by $Q$ and $R$. The isolines of probability in the $Q-R$ plane express intimate features of the turbulent flow geometry, has a nontrivial teardrop shaped pattern documented in the literature~\cite{elsinga2010universal,chertkov1999lagrangian}. Different parts of the $Q-R$ plane are associated with different structures of the flow. Thus lower right corner (negative $Q$ and $R$), which has higher probability than other quadrants, corresponds to a pancake type of structure (two expanding directions, one contracting) with the direction of rotation (vorticity) aligned with the second eigenvector of the stress. A more nuanced analysis of the $Q-R$ PDF at various turbulence scales is possible by coarse-graining the velocity gradient tensor over a region $\Gamma$ with a characteristic scale $r$, which lies in the inertial range. The region $\Gamma$ can be thought of as a local correlation volume of velocity gradient which is constructed around a pre-defined scale $r$. Therefore, this scale dictates the size of the correlation volume $\Gamma$ and hence the Lagrangian dynamics contained in it correspond to the scale $r$. This allows us to perform coarse-graining of the flow statistics, such that $Q-R$ PDFs can be studied for various turbulence scales. A direct advantage of using this approach for machine learning based predictive models is that their accuracy/deficiency at various turbulence scales can be studied in significant physical detail, as done in this work. As a result, the tea-drop shape of the probability isoline becomes more prominent with decrease of the coarse-graining scale. 


\bibliography{main}

\bibliographystyle{ieeetr}

\end{document}